\documentclass[aps,prl,reprint,superscriptaddress]{revtex4-2}

\usepackage{graphicx}
\usepackage{xcolor}
\usepackage{dcolumn}
\usepackage{bm, bbm, siunitx, gensymb, xfrac, physics}
\usepackage{hyperref}
\usepackage{booktabs, multirow}

\hypersetup{
    colorlinks=true,
    urlcolor= blue,
    citecolor=blue,
    linkcolor= blue,
    bookmarksopen=false,
    }
    
\newcounter{para}
\newcommand\mypara{\par\refstepcounter{para}\noindent \textbf{\thepara}\indent}

\begin{document}

\title{Topological phononic logic}
\author{Harris Pirie}
\affiliation{Department of Physics, Harvard University, Cambridge, MA, 02138, USA}
\author{Shuvom Sadhuka}
\affiliation{School of Engineering and Applied Science, Harvard University, Cambridge, MA, 02138, USA}
\author{Jennifer Wang}
\affiliation{School of Engineering and Applied Science, Harvard University, Cambridge, MA, 02138, USA}
\affiliation{Department of Physics, Wellesley College, Wellesley MA, 02481, USA}
\author{Radu Andrei}
\affiliation{Department of Physics, Harvard University, Cambridge, MA, 02138, USA}
\author{Jennifer E. Hoffman}
\email{jhoffman@physics.harvard.edu}
\affiliation{Department of Physics, Harvard University, Cambridge, MA, 02138, USA}
\affiliation{School of Engineering and Applied Science, Harvard University, Cambridge, MA, 02138, USA}
\date{\today}

\begin{abstract}
Topological metamaterials have robust properties engineered from their macroscopic arrangement, rather than their microscopic constituency. They can be designed by starting from Dirac metamaterials with either symmetry-enforced or accidental degeneracy. The latter case provides greater flexibility in the design of topological switches, waveguides, and cloaking devices, because a large number of tuning parameters can be used to break the degeneracy and induce a topological phase. However, the design of a topological logic element---a switch that can be controlled by the output of a separate switch---remains elusive. Here we numerically demonstrate a topological logic gate for ultrasound by exploiting the large phase space of accidental degeneracies in a honeycomb lattice.  We find that a degeneracy can be broken by six physical parameters, and we show how to tune these parameters to create a phononic switch that transitions between a topological waveguide and a trivial insulator by ultrasonic heating.  Our design scheme is directly applicable to photonic crystals and may guide the design of future electronic topological transistors. 
\end{abstract}

\maketitle

\mypara
Topological insulators were first conceived as quantum electronic materials with an insulating bulk and conducting surface Dirac states, allowing for dissipationless charge and spin transport along their boundaries. Their central principle---the inversion of energy bands---is also present in many classical lattice systems, inspiring the design of photonic \cite{HaldanePhysRevLett2008, KhanikaevNatMater2013, LuNatPhotonics2014}, phononic \cite{GeNatlSciRev2018}, and mechanical metamaterials \cite{KaneNatPhys2014, SusstrunkScience2015, HuberNatPhys2016} with topologically protected transport. These classical systems provide a platform to test ideas in topological band theory, because they are more tangibly understood than their quantum counterparts, and their governing wave equations can be solved exactly. Their robust properties have been used in many promising applications including  zero- and negative-refractive-index materials \cite{LiuApplPhysLett2012, LiApplPhysLett2014, DuboisNatCommun2017, LiberalNatPhotonics2017, HeNature2018}, cloaking \cite{HaoApplPhysLett2010, HuangNatMater2011}, and protected waveguides for sound and light that outperform non-topological alternatives \cite{MousaviNatCommun2015, HeNatPhys2016, WeiPhysRevB2017}. A key remaining challenge is to control the topological phase in a way that allows waveguides to toggle one another, paving the way towards topological logic circuits with greater efficiency than current CMOS technology \cite{CheckelskyNatPhys2012, WrayNatPhys2012, CTW18}.

\mypara
A general design approach to achieve the band inversion that defines a topological metamaterial is to start from a bulk Dirac state, then intentionally break the Dirac-point degeneracy to open a negative gap. This approach can be broadly divided into two methods.  The first method starts from a symmetry-enforced Dirac state, such as the $K$-point Dirac cone in graphene-like honeycomb or triangular metamaterials, then opens a gap by breaking a symmetry of the system.  In systems with broken time-reversal ($\mathcal{T}$) symmetry \cite{WangNature2009, FleuryScience2014, YangPhysRevLett2015, WangPhysRevLett2015, NashProcNatlAcadSci2015}, the resultant topological phase is analogous to the quantum Hall effect, while those with broken inversion symmetry \cite{WuPhysRevLett2015, ZhangPhysRevLett2017, ZhangPhysRevB2017, DengPhysRevB2017, YangPhysRevLett2018, XiaPhysRevB2017} can realize an analog of the quantum spin Hall effect.  However, there is limited flexibility in the design of these topological phases, as they can be tuned only by a symmetry-breaking operation. On the other hand, the second method searches for the accidental degeneracy of three \cite{HuangNatMater2011, MeiPhysRevB2012, LiuApplPhysLett2012} or four \cite{SakodaOptExpress2012, LiApplPhysLett2014, ChenSciRep2015} bands, producing a Dirac-like cone or double Dirac cones, respectively. This method gives access to a far larger set of topological phases because the accidental degeneracy can be broken by several more-accessible tuning parameters while retaining inversion and $\mathcal{T}$ symmetry. Despite the utility and flexibility of this second method, the complete space of all topological phases has yet to be mapped for any accidental degeneracy. 

\mypara
We start from a particular accidental bulk Dirac-point degeneracy that gives rise to a topological state analogous to a quantum spin Hall system. In a quantum spin Hall system, the protection of the Dirac point is a consequence of the spin-$\sfrac{1}{2}$ nature of electrons. Specifically, because $\mathcal{T}^2=-1$ for spin-$\sfrac{1}{2}$ states, Kramers theorem requires a degeneracy at all $\mathcal{T}$-invariant points of the Brillouin zone. However, spin-0 phononic and spin-1 photonic systems both have $\mathcal{T}^2=+1$, so Kramers theorem does not apply.  Instead, designs typically rely on mode hybridization to form a pseudospin-$\sfrac{1}{2}$ subsystem, for example with the transverse electric and magnetic polarizations of light \cite{KhanikaevNatMater2013}.  But transverse shear modes are not available in airborne acoustics, so finding an analogy of Kramers theorem is challenging. In 2012, Sakoda \cite{SakodaOptExpress2012} addressed this issue and constructed a pseudospin-$\sfrac{1}{2}$ system using the discrete symmetries of a triangular lattice, which was adapted to longitudinal acoustic modes shortly thereafter \cite{LiApplPhysLett2014, ChenSciRep2015, MeiSciRep2016}, and subsequently demonstrated experimentally \cite{HeNatPhys2016}. In this scheme, a lattice with $C_{6v}$ symmetry generates an accidental degeneracy at the $\Gamma$ point between doubly degenerate $E_1$ and $E_2$ modes that transform as $(x,y)$ and $(xy, x^2-y^2)$, denoted $(p_x, p_y)$ and $(d_{xy}, d_{x^2-y^2})$, respectively. These doubly degenerate modes allow the formation of a pseudospin-$1/2$ basis, with corresponding eigenstates $p_\pm = (p_x \pm i p_y)/\sqrt{2}$ and $d_{\pm} = (d_{x^2-y^2} \pm i d_{xy})/\sqrt{2}$. The accidental degeneracy between the $p_\pm$ and $d_\pm$ subsystems can be lifted without breaking $C_{6v}$ symmetry, resulting in a topological phase with helical edge modes protected by a pseudo-$\mathcal{T}$ symmetry, analogous to the quantum spin Hall state \cite{HeNatPhys2016, WuPhysRevLett2015}.

\mypara
Here we numerically investigate the topological phase space for a $\Gamma$-point accidental degeneracy in a phononic honeycomb lattice using commercial finite-element modeling software {\sc comsol multiphysics}.   We find a manifold of system configurations that host a bulk accidental double Dirac cone, and we demonstrate that a topological phase can be induced by gapping the Dirac node with six independent physical parameters, which collapse into a three-dimensional (3D) phase space.  This vast phase space guides the design of three topological circuit elements: a static-geometry waveguide, an externally switchable device, and a universal logic gate. While we illustrate each element using phononic metamaterials, the same design principles apply to electronic and photonic metamaterials.

\begin{figure}
	\includegraphics[width=0.48\textwidth]{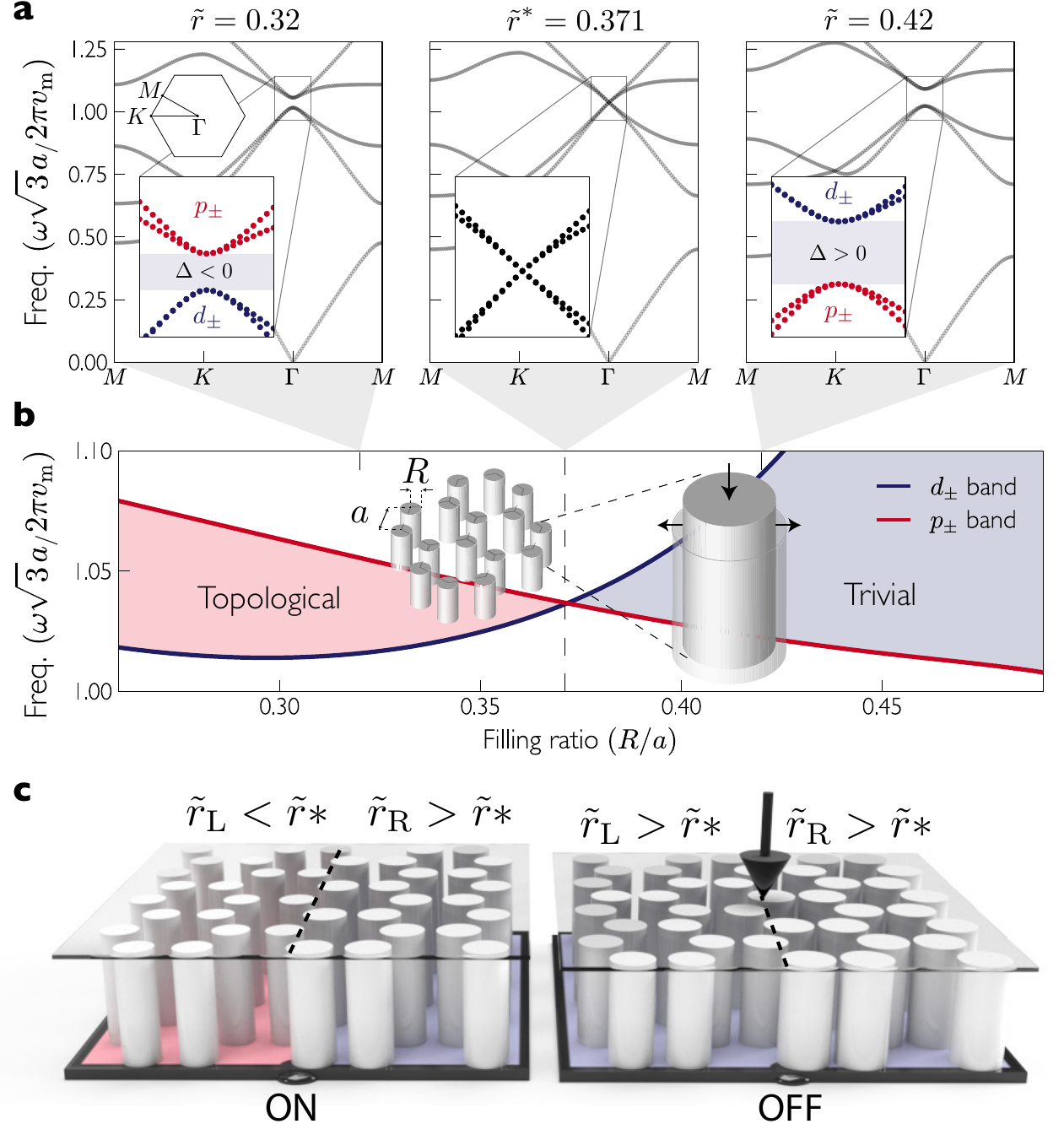}
	\caption{\label{fig:1} 
	{\bf An externally controlled topological switch for sound.}
    ({\bf a}) The phononic band structure for a honeycomb lattice of steel pillars in water passes through an accidental degeneracy as the radius of the pillars is varied.  This degeneracy is between $p_\pm$ bands (red) and  $d_\pm$ bands (blue), and occurs at the critical filling ratio of $\tilde{r}^* \equiv R^*/a = 0.371$ (middle panel).  As the filling ratio is tuned away from this value, a positive (right) or negative (left) band gap opens, leading to a topological phase transition. 
    ({\bf b}) This transition can be clearly seen by tracking the $\Gamma$-point eigenvalues as $\tilde{r}$ is tuned. 
    ({\bf c}) A topological waveguide is made by placing two lattices with $\tilde{r_L}<\tilde{r}^*$ and $\tilde{r_R}>\tilde{r}^*$  next to each other (left panel).  When the pillars are compressed vertically, their radius expands such that both sides of the waveguide become trivial insulators (right panel).  This device is a topological switch for sound that turns `off' when compressed. 
    }
\end{figure}

\mypara
A static-geometry topological waveguide is formed at the interface between a lattice with normally ordered bands and one with inverted bands. This type of waveguide was already demonstrated using two hexagonal phononic crystals of steel pillars in a fluid medium with different filling ratios, $\tilde{r}=R/a$ \cite{ChenSciRep2015, HeNatPhys2016}, where $R$ and $a$ are the radius and spacing of the pillars, respectively (see inset to Fig.~\hyperref[fig:1]{1(b)}).  When the filling ratio is large, the band structure around the $\Gamma$ point contains doubly degenerate $p_\pm$ modes separated from $d_\pm$ modes by a positive energy gap, $\Delta > 0$, as shown in Fig.~\hyperref[fig:1]{1(a)}. At the critical filling ratio for a steel/water system, $\tilde{r}^* = 0.371$, the four modes become accidentally degenerate and the bulk metamaterial hosts double Dirac cones. Below critical filling, the $p_\pm$ modes have higher energy than the $d_\pm$ modes, and the band structure contains a negative energy gap, $\Delta < 0$.  Topologically protected edge modes are confined to the interface between a positive- and negative-gapped material, allowing the design of topological waveguides that are pseudospin polarized and immune to defects including cavities, bends, and lattice disorder \cite{HeNatPhys2016}. 

\mypara
Our first advance is a new mechanism to create an externally switchable topological waveguide for sound, providing a simple alternative to the existing schemes \cite{SZH17, ZhangPhysRevAppl2018, XJS18}. In general, a topological switch hosts robust transport when `on', but is a trivial insulator when `off'. It requires a tuning mechanism capable of changing the sign of the band gap on the topological side, while leaving the trivial side unchanged. We found that an external vertical compression/extension can induce this behavior, as it alters the radius of pillars, which can toggle the topological phase (see inset to Fig.~\hyperref[fig:1]{1(b)}). For materials with a positive Poisson's ratio, a topological waveguide naturally switches `off' when compressed, once the filling ratio of its topological side increases beyond $\tilde{r}^*$, as shown in Fig.~\hyperref[fig:1]{1(c)}.  In practice, rubber pillars are ideal for this application as they are far more stretchable than metal pillars, and have a higher Poisson's ratio (see Supplementary Information). Advancing beyond static-geometry topological waveguides \cite{MousaviNatCommun2015, HeNatPhys2016, WeiPhysRevB2017, DengPhysRevB2017, ZhangPhysRevB2017, XiaPhysRevB2017, YangPhysRevLett2018}, this type of switch could be used to control passive acoustic isolation systems, but the output of one switch cannot sustain the macroscopic stretch required to activate a second, similar switch.   

\begin{figure}
	\includegraphics[width=0.48\textwidth]{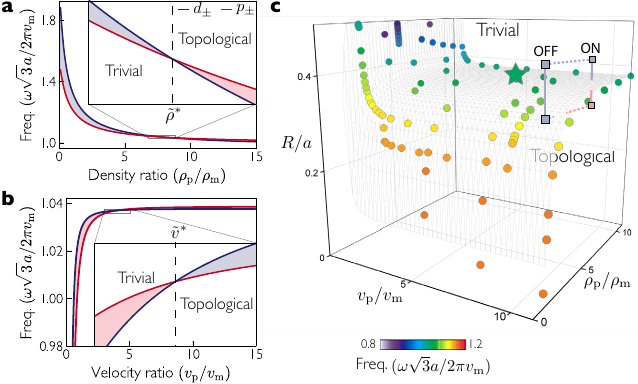}
	\caption{\label{fig:2} 
	{\bf Topological phase space for a honeycomb phononic lattice.}
	An accidental degeneracy between the $p_\pm$ and $d_\pm$ modes in a steel/water system (green star in (c)) can be broken by tuning the ratio of ({\bf a}) density while holding speed of sound and radius fixed; or ({\bf b}) speed of sound while holding radius and density fixed. 
	({\bf c}) Each accidental degeneracy is a point in $(\tilde{v},\tilde{\rho},\tilde{r})$ space, colored according to its crossing frequency (e.g.~the steel/water system has a crossing frequency of 90 kHz for $a=1$ cm).  Together, they separate phase space into a topological and a trivial region. 
	A topological waveguide pairs configurations from different sides of this surface (see solid line labelled ON), provided their bulk spectral gaps overlap. Transmission through it can be switched `off' by tuning to two configurations that occur on the same side of the surface (see path labelled OFF). 
    }
\end{figure}

\mypara
Our second, more significant advance is to design a phononically controlled acoustic switch---i.e.~a topological logic element. Like electronic field-effect transistors, these switches may be connected together to form circuits. Here we rely explicitly on the flexibility granted by the large phase space of accidental degeneracies in a honeycomb metamaterial. In general, an accidental band degeneracy can be lifted by tuning any lattice parameter, as it is not protected by symmetry.  The relevant parameters in a phononic lattice define the acoustic wave equation,
\begin{align}
	\label{eqn:wave}
	\div{\qty[\frac{1}{\rho_r(\vb{r})} \grad p(\vb{r})]} 
	    = -\frac{\omega^2}{v_\mathrm{m}^2} \cdot \frac{p(\vb{r})}{v_r^2(\vb{r}) \rho_r(\vb{r})}
\end{align}
where $p$ is the pressure, $\omega$ is the eigenfrequency, and $\rho_r(\vb{r})=\rho(\vb{r})/\rho_{\mathrm{m}}$ and $v_r(\vb{r})=v(\vb{r})/v_{\mathrm{m}}$ are the relative density and speed of sound, respectively.  In total, there are six physical parameters that can tune the resulting eigenspectrum: $R$, $a$, $\rho_{\mathrm{p}}$, $\rho_{\mathrm{m}}$, $v_{\mathrm{p}}$, and $v_{\mathrm{m}}$, where the subscript refers to pillars or medium. First note that uniformly scaling $\rho_{\mathrm{p}}$ and $\rho_{\mathrm{m}}$ produces no change. Second, uniformly scaling $v_{\mathrm{p}}$ and $v_{\mathrm{m}}$ scales all eigenfrequencies of Eq.~\ref{eqn:wave}, but does not shift eigenfrequencies relative to one another, and therefore cannot alter the topological phase. We take this scaling into account by adopting dimensionless units for frequency, $\tilde{\omega} = \omega \sqrt{3}a / 2\pi v_{\mathrm{m}}$. In fact, the frequency-normalized band structure depends only on three dimensionless ratios: $\tilde{r} = R/a$, $\tilde{v} = v_{\mathrm{p}}/v_{\mathrm{m}}$ and $\tilde{\rho} = \rho_{\mathrm{p}}/\rho_{\mathrm{m}}$.  In the example system of steel pillars in water, we find that varying either $\tilde{v}$ or $\tilde{\rho}$ lifts the accidental degeneracy and can open a negative gap (Fig.~\hyperref[fig:2]{2(a-b)}). More generally, varying any combination of lattice parameters along a path in $(\tilde{v},\tilde{\rho},\tilde{r})$ space that connects the topological phase to the trivial phase must pass through an accidental degeneracy.  Consequently, there exists a surface in  $(\tilde{v},\tilde{\rho},\tilde{r})$ space that separates the topological phase from the trivial phase, on which there is accidental degeneracy between $p_\pm$ and $d_\pm$ modes and a bulk double Dirac cone.  We numerically calculated the shape of this surface, shown in Fig.~\hyperref[fig:2]{2(c)}, by recording the accidental crossing point in an $\tilde{r}$ sweep for a discrete set of $(\tilde{v},\tilde{\rho})$ points, at fixed $({v_{\mathrm{m}}, \rho_{\mathrm{m}}})$.

\mypara
A key challenge in designing a topological switch is to preserve overlapping bulk spectral gaps before and after switching. For example, in the sweep shown in Fig.~\hyperref[fig:2]{2(a)}, increasing $\tilde{\rho}$ causes both $p_\pm$ and $d_\pm$ modes to decrease in frequency, leading to a band inversion because the $d_\pm$ modes decrease faster than the $p_\pm$ modes. Yet, this tuning parameter alone cannot be used to design a topological waveguide because at any frequency there are bulk modes in one of the two sides that mask the edge states, unlike Fig.~\hyperref[fig:1]{1(b)}. The same accidental degeneracy can be broken by varying $\tilde{v}$, which causes both $p_\pm$ and $d_\pm$ modes to increase in frequency (Fig.~\hyperref[fig:2]{2(b)}), again precluding a usefully overlapping gap.  However, an overlapping bulk gap may occur when tuning a combination of $\tilde{v}$ and $\tilde{\rho}$, for instance, in a waveguide between two sets of pillars with different materials but the same radius.  In general, each accidental degeneracy on the surface in Fig.~\hyperref[fig:2]{2(c)} can be used to construct a practical waveguide for a parameter sweep through some solid angle in $(\tilde{v},\tilde{\rho},\tilde{r})$ space. Schematically, such a waveguide combines two points in parameter space connected by a path that punctures the surface in Fig.~\hyperref[fig:2]{2(c)}.  Consequently, a topological switch combines four points in phase space, with three above the surface (trivial) and one below (topological), e.g.~the square points in Fig.~\hyperref[fig:2]{2(c)}.  Furthermore, a useful switch requires the bulk to remain gapped and overlapping on all four $(\tilde{v},\tilde{\rho},\tilde{r})$ trajectories that connect these points, except where they pass through the surface. 

\begin{figure}
	\includegraphics[width=0.48\textwidth]{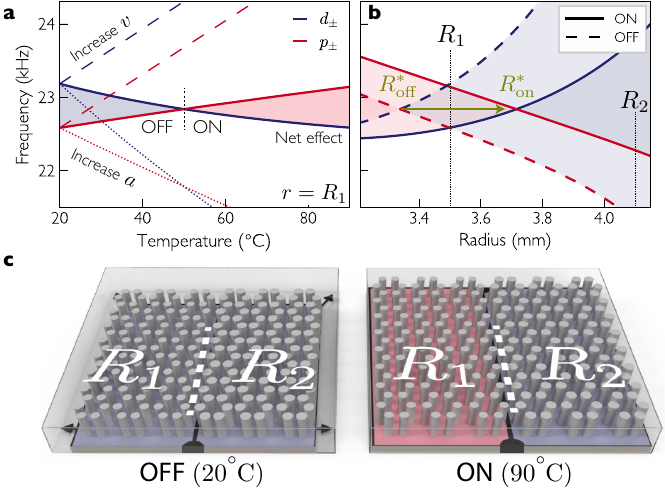}
	\caption{\label{fig:3} 
	{\bf Designing a temperature-controlled topological switch.} We consider a honeycomb lattice of steel pillars ($R_1=3.5$ mm, $R_2 = 4.1$ mm, $a=8.5$ mm) anchored to a high-thermal-expansion base plate, in an air medium within a sealed fixed-size box.
    ({\bf a}) Heating this system has two main competing effects: eigenfrequencies are increased by raising the speed of sound in air (dashed lines), but decreased as the base plate thermally expands (dotted lines). The latter effect also tunes $\tilde{r}$ to induce a band inversion. These two effects can be balanced by correctly choosing the thermal expansion coefficient of the base plate (here $1.61\times 10^{-3}$ K$^{-1}$), providing a temperature-tunable topological phase transition with an overlapping spectral gap (solid lines).
    ({\bf b}) A topological switch combines two sizes of pillars: one side transitions from trivial to topological as the switch is heated $(R_1)$, while the other remains trivial throughout $(R_2)$.
    ({\bf c}) Unlike the switch design in Fig.~\hyperref[fig:1]{1 (c)}, which is triggered by tuning $R$ at fixed $a$, this switch is turned `on' by increasing $a$ at fixed $R$, and can be actuated by phonon-delivered heat. 
    }
\end{figure}

\begin{figure}
	\includegraphics[width=0.48\textwidth]{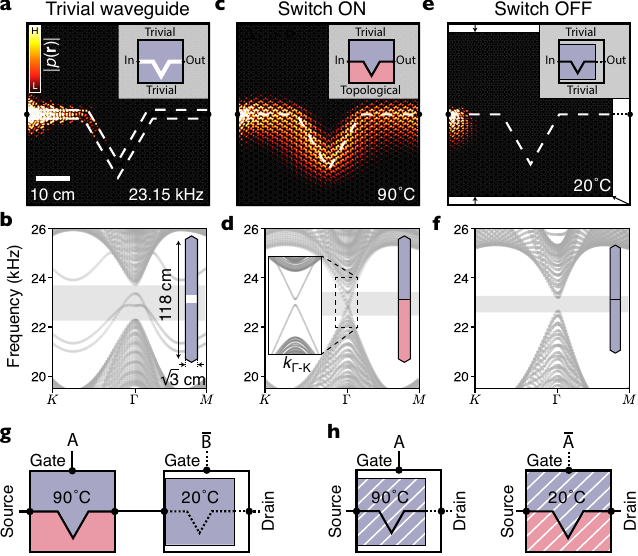}
	\caption{\label{fig:4} 
	{\bf Topological logic gates with ultrasound.}
    ({\bf a}) Transmission through a trivial waveguide, like this channel in an insulating steel/air phononic crystal, is disrupted by disorder and bends. 
    ({\bf b}) The band structure of the corresponding supercell (see inset) contains a bulk gap with trivial edge states. 
    ({\bf c}) In contrast, a topological waveguide with the same parameters as in Fig.~3, allows robust transport regardless of channel geometry; it can be used as the `on' state of a topological switch. 
    ({\bf d}) Its band structure hosts protected, Dirac-like edge states (inset) due to a negative bulk gap on one side.
    ({\bf e}) When the system is cooled, it contracts and both sides become trivial insulators, preventing transmission.
    ({\bf f}) Excitations at frequencies within the gap decay exponentially as they enter the device. 
    ({\bf g}) A topological {\sc and} gate, constructed from two switches in series, requires both control signals (A and B) to be high to register an output.
    ({\bf h}) A topological {\sc not} gate uses a base plate with a negative thermal expansion coefficient; it contracts to turn `off' when heated (left), and expands to turn `on' when cooled (right).
    }
\end{figure}

\mypara 
To enable the output of one switch to control the next, our design for a phononically-controlled topological switch uses a temperature increase delivered by ultrasonic phonons as its tuning mechanism. Each switch contains a honeycomb lattice of steel pillars connecting the source and drain terminals, attached to a base plate made from a second material, in an air-tight container, as shown in Fig.~\hyperref[fig:3]{3(c)}. The temperature increase needed to toggle the switch is provided by a thermoacoustic converter connected to a third terminal (labelled `gate' in Fig.~\hyperref[fig:4]{4(g-h)}, see Supplementary Information). The primary effect of heating the device is to change the speed of sound in the medium, which typically increases all eigenfrequencies of the system (see dashed lines in Fig.~\hyperref[fig:3]{3(a)}).  Second, heating  causes thermal expansion of the materials, increasing both $R$ and $a$, though not necessarily equally. If the base plate and pillar materials are selected such that $a$ increases faster than $R$, the net result is to reduce all eigenfrequencies of the system and induce a band inversion (dotted lines in  Fig.~\hyperref[fig:3]{3(a)}).  Finally, heating alters the density of the air and the materials, which has been taken into account, but is insignificant.  The first two effects can be balanced to maintain a bulk gap throughout the switching process, by fixing the ratio $v_\text{m}/a$ that appears in the eigenfrequency of Eq.~\hyperref[eqn:wave]{1}.  Because the base plate expands linearly with temperature, we seek a medium where $v_\text{m}$ also increases linearly.  For an ideal gas at temperature $T$, $v_\mathrm{m}$ increases as $\sqrt{T}$, but the trend is almost linear near room temperature; as such, air is a suitable medium.  Consequently, to keep $v_\text{m}/a$ fixed as the temperature increases from $T_i$ to $T_f$, we seek a base material with a coefficient of thermal expansion given by $\alpha=1/(T_i + \sqrt{T_iT_f})$. For the proof-of-principle switch shown in Fig.~\hyperref[fig:3]{3(b)}, the required base-plate thermal expansion coefficient is $1.61 \times 10^{-3}$ K$^{-1}$, which is within the range achievable by origami metamaterials \cite{BoattiAdvMater2017}.  Alternatively, we empirically demonstrated an even larger effective thermal expansion coefficient by thermally actuating using the shape-memory alloy Nitinol (see Supplementary Information).

\mypara
The advantage of a topological phononic switch can be seen from the finite-sized calculations in Fig.~\hyperref[fig:4]{4}.  Unlike a trivial waveguide, which experiences significant losses induced by disorder and bends (Fig.~\hyperref[fig:4]{4(a-b)}), the topological switch acts as a robust pseudospin-dependent waveguide when `on' due to a Dirac cone between the two sides (Fig.~\hyperref[fig:4]{4(c-d)}). As it is cooled, the pillars contract around the input terminal; both sides become trivially insulating and block transmission, turning the switch `off' (Fig.~\hyperref[fig:4]{4(e-f)}). Our topological switch is stable against mild temperature changes provided its operational frequency remains within the spectral gap (see Fig.~\hyperref[fig:3]{3(a)}). Such temperature variations alter only the localization of the edge states, not their presence or absence (see Supplementary Information).

\mypara
Our proof-of-principle temperature-controlled phononic switches can be linked to form a universal {\sc nand} gate with two main segments. First, we design a topological {\sc and} gate by connecting two switches in series (Fig.~\hyperref[fig:4]{4(g)}). This device requires both control signals (A and B) to be `on' to heat each switch and allow information to propagate (see Supplementary Information). Second, to design a topological {\sc not} gate, we utilize a base plate material that has a negative coefficient of thermal expansion; that is, it shrinks when heated. At room temperature (control is `off'), the {\sc not} gate is a topological waveguide that transmits information, but when the control is `on', the device heats and shrinks, transitioning to a trivial insulator.  To maintain an overlapping bulk gap throughout this transition, we require a medium where the speed of sound decreases with increasing temperature, a behavior commonly observed in oils \cite{OliveiraJPhysConfSer2016}. Specifically, a device using steel pillars in sunflower oil requires a coefficient of thermal expansion of $-2.0\times 10^{-3}$ K$^{-1}$ to keep the ratio $v_\text{m}/a$ fixed, a value recently demonstrated \cite{BoattiAdvMater2017}.  

\mypara
The design of topological metamaterials based on a broken accidental degeneracy is extremely versatile due to the large number of tuning parameters available.  Specifically, for a phononic honeycomb lattice, the topological phase can be tuned by six independent parameters, which collapse onto a 3D phase space. This phase space guided a proof-of-principle design for a phononically controlled topological switch, the building block of an acoustic logic gate. The macroscopic size and moderate speed of our device makes it an ideal tool for teaching and understanding topological materials. More importantly, the same design process can be followed for piezoelectric materials at mesoscopic length scales, enabling switchable control of topologically protected surface-acoustic waves for integrated phononics \cite{YSN16}. Finally, our approach directly applies to optical systems under a simple mapping of variables \cite{MeiPhysRevB2012}, or to nano-structured quantum materials \cite{PoliniNatNano2013}, providing a new direction for developing a field-effect topological transistor. 

\begin{acknowledgments}
We thank Katia Bertoldi, Barbara Dorritie, Jason Hoffman, David Lee, Ciar\'{a}n O'Neill, Ben November, Pai Wang, and Jack Zhang for helpful conversations.  This work was supported by the Science and Technology Center for Integrated Quantum Materials under the National Science Foundation grant No.~DMR-1231319. 
\end{acknowledgments}

\bibliography{refs}

\begin{thebibliography}{42}%
\makeatletter
\providecommand \@ifxundefined [1]{%
 \@ifx{#1\undefined}
}%
\providecommand \@ifnum [1]{%
 \ifnum #1\expandafter \@firstoftwo
 \else \expandafter \@secondoftwo
 \fi
}%
\providecommand \@ifx [1]{%
 \ifx #1\expandafter \@firstoftwo
 \else \expandafter \@secondoftwo
 \fi
}%
\providecommand \natexlab [1]{#1}%
\providecommand \enquote  [1]{``#1''}%
\providecommand \bibnamefont  [1]{#1}%
\providecommand \bibfnamefont [1]{#1}%
\providecommand \citenamefont [1]{#1}%
\providecommand \href@noop [0]{\@secondoftwo}%
\providecommand \href [0]{\begingroup \@sanitize@url \@href}%
\providecommand \@href[1]{\@@startlink{#1}\@@href}%
\providecommand \@@href[1]{\endgroup#1\@@endlink}%
\providecommand \@sanitize@url [0]{\catcode `\\12\catcode `\$12\catcode
  `\&12\catcode `\#12\catcode `\^12\catcode `\_12\catcode `\%12\relax}%
\providecommand \@@startlink[1]{}%
\providecommand \@@endlink[0]{}%
\providecommand \url  [0]{\begingroup\@sanitize@url \@url }%
\providecommand \@url [1]{\endgroup\@href {#1}{\urlprefix }}%
\providecommand \urlprefix  [0]{URL }%
\providecommand \Eprint [0]{\href }%
\providecommand \doibase [0]{https://doi.org/}%
\providecommand \selectlanguage [0]{\@gobble}%
\providecommand \bibinfo  [0]{\@secondoftwo}%
\providecommand \bibfield  [0]{\@secondoftwo}%
\providecommand \translation [1]{[#1]}%
\providecommand \BibitemOpen [0]{}%
\providecommand \bibitemStop [0]{}%
\providecommand \bibitemNoStop [0]{.\EOS\space}%
\providecommand \EOS [0]{\spacefactor3000\relax}%
\providecommand \BibitemShut  [1]{\csname bibitem#1\endcsname}%
\let\auto@bib@innerbib\@empty
\bibitem [{\citenamefont {Haldane}\ and\ \citenamefont
  {Raghu}(2008)}]{HaldanePhysRevLett2008}%
  \BibitemOpen
  \bibfield  {author} {\bibinfo {author} {\bibfnamefont {F.~D.~M.}\
  \bibnamefont {Haldane}}\ and\ \bibinfo {author} {\bibfnamefont
  {S.}~\bibnamefont {Raghu}},\ }\bibfield  {title} {\bibinfo {title} {Possible
  {{Realization}} of {{Directional Optical Waveguides}} in {{Photonic
  Crystals}} with {{Broken Time}}-{{Reversal Symmetry}}},\ }\href
  {https://doi.org/10.1103/PhysRevLett.100.013904} {\bibfield  {journal}
  {\bibinfo  {journal} {Physical Review Letters}\ }\textbf {\bibinfo {volume}
  {100}},\ \bibinfo {pages} {013904} (\bibinfo {year} {2008})}\BibitemShut
  {NoStop}%
\bibitem [{\citenamefont {Khanikaev}\ \emph {et~al.}(2013)\citenamefont
  {Khanikaev}, \citenamefont {Mousavi}, \citenamefont {Tse}, \citenamefont
  {Kargarian}, \citenamefont {MacDonald},\ and\ \citenamefont
  {Shvets}}]{KhanikaevNatMater2013}%
  \BibitemOpen
  \bibfield  {author} {\bibinfo {author} {\bibfnamefont {A.~B.}\ \bibnamefont
  {Khanikaev}}, \bibinfo {author} {\bibfnamefont {S.~H.}\ \bibnamefont
  {Mousavi}}, \bibinfo {author} {\bibfnamefont {W.-K.}\ \bibnamefont {Tse}},
  \bibinfo {author} {\bibfnamefont {M.}~\bibnamefont {Kargarian}}, \bibinfo
  {author} {\bibfnamefont {A.~H.}\ \bibnamefont {MacDonald}},\ and\ \bibinfo
  {author} {\bibfnamefont {G.}~\bibnamefont {Shvets}},\ }\bibfield  {title}
  {\bibinfo {title} {Photonic topological insulators},\ }\href
  {https://doi.org/10.1038/nmat3520} {\bibfield  {journal} {\bibinfo  {journal}
  {Nature Materials}\ }\textbf {\bibinfo {volume} {12}},\ \bibinfo {pages}
  {233} (\bibinfo {year} {2013})}\BibitemShut {NoStop}%
\bibitem [{\citenamefont {Lu}\ \emph {et~al.}(2014)\citenamefont {Lu},
  \citenamefont {Joannopoulos},\ and\ \citenamefont {Solja{\v
  c}i\'c}}]{LuNatPhotonics2014}%
  \BibitemOpen
  \bibfield  {author} {\bibinfo {author} {\bibfnamefont {L.}~\bibnamefont
  {Lu}}, \bibinfo {author} {\bibfnamefont {J.~D.}\ \bibnamefont
  {Joannopoulos}},\ and\ \bibinfo {author} {\bibfnamefont {M.}~\bibnamefont
  {Solja{\v c}i\'c}},\ }\bibfield  {title} {\bibinfo {title} {Topological
  photonics},\ }\href {https://doi.org/10.1038/nphoton.2014.248} {\bibfield
  {journal} {\bibinfo  {journal} {Nature Photonics}\ }\textbf {\bibinfo
  {volume} {8}},\ \bibinfo {pages} {821} (\bibinfo {year} {2014})}\BibitemShut
  {NoStop}%
\bibitem [{\citenamefont {Ge}\ \emph {et~al.}(2018)\citenamefont {Ge},
  \citenamefont {Yang}, \citenamefont {Ma}, \citenamefont {Lu}, \citenamefont
  {Chen}, \citenamefont {Fang},\ and\ \citenamefont
  {Sheng}}]{GeNatlSciRev2018}%
  \BibitemOpen
  \bibfield  {author} {\bibinfo {author} {\bibfnamefont {H.}~\bibnamefont
  {Ge}}, \bibinfo {author} {\bibfnamefont {M.}~\bibnamefont {Yang}}, \bibinfo
  {author} {\bibfnamefont {C.}~\bibnamefont {Ma}}, \bibinfo {author}
  {\bibfnamefont {M.-H.}\ \bibnamefont {Lu}}, \bibinfo {author} {\bibfnamefont
  {Y.-F.}\ \bibnamefont {Chen}}, \bibinfo {author} {\bibfnamefont
  {N.}~\bibnamefont {Fang}},\ and\ \bibinfo {author} {\bibfnamefont
  {P.}~\bibnamefont {Sheng}},\ }\bibfield  {title} {\bibinfo {title} {Breaking
  the barriers: Advances in acoustic functional materials},\ }\href
  {https://doi.org/10.1093/nsr/nwx154} {\bibfield  {journal} {\bibinfo
  {journal} {National Science Review}\ }\textbf {\bibinfo {volume} {5}},\
  \bibinfo {pages} {159} (\bibinfo {year} {2018})}\BibitemShut {NoStop}%
\bibitem [{\citenamefont {Kane}\ and\ \citenamefont
  {Lubensky}(2014)}]{KaneNatPhys2014}%
  \BibitemOpen
  \bibfield  {author} {\bibinfo {author} {\bibfnamefont {C.~L.}\ \bibnamefont
  {Kane}}\ and\ \bibinfo {author} {\bibfnamefont {T.~C.}\ \bibnamefont
  {Lubensky}},\ }\bibfield  {title} {\bibinfo {title} {Topological boundary
  modes in isostatic lattices},\ }\href {https://doi.org/10.1038/nphys2835}
  {\bibfield  {journal} {\bibinfo  {journal} {Nature Physics}\ }\textbf
  {\bibinfo {volume} {10}},\ \bibinfo {pages} {39} (\bibinfo {year}
  {2014})}\BibitemShut {NoStop}%
\bibitem [{\citenamefont {Susstrunk}\ and\ \citenamefont
  {Huber}(2015)}]{SusstrunkScience2015}%
  \BibitemOpen
  \bibfield  {author} {\bibinfo {author} {\bibfnamefont {R.}~\bibnamefont
  {Susstrunk}}\ and\ \bibinfo {author} {\bibfnamefont {S.~D.}\ \bibnamefont
  {Huber}},\ }\bibfield  {title} {\bibinfo {title} {Observation of phononic
  helical edge states in a mechanical topological insulator},\ }\href
  {https://doi.org/10.1126/science.aab0239} {\bibfield  {journal} {\bibinfo
  {journal} {Science}\ }\textbf {\bibinfo {volume} {349}},\ \bibinfo {pages}
  {47} (\bibinfo {year} {2015})}\BibitemShut {NoStop}%
\bibitem [{\citenamefont {Huber}(2016)}]{HuberNatPhys2016}%
  \BibitemOpen
  \bibfield  {author} {\bibinfo {author} {\bibfnamefont {S.~D.}\ \bibnamefont
  {Huber}},\ }\bibfield  {title} {\bibinfo {title} {Topological mechanics},\
  }\href {https://doi.org/10.1038/nphys3801} {\bibfield  {journal} {\bibinfo
  {journal} {Nature Physics}\ }\textbf {\bibinfo {volume} {12}},\ \bibinfo
  {pages} {621} (\bibinfo {year} {2016})}\BibitemShut {NoStop}%
\bibitem [{\citenamefont {Liu}\ \emph {et~al.}(2012)\citenamefont {Liu},
  \citenamefont {Huang},\ and\ \citenamefont {Chan}}]{LiuApplPhysLett2012}%
  \BibitemOpen
  \bibfield  {author} {\bibinfo {author} {\bibfnamefont {F.}~\bibnamefont
  {Liu}}, \bibinfo {author} {\bibfnamefont {X.}~\bibnamefont {Huang}},\ and\
  \bibinfo {author} {\bibfnamefont {C.~T.}\ \bibnamefont {Chan}},\ }\bibfield
  {title} {\bibinfo {title} {Dirac cones at $\vec{k}=0$ in acoustic crystals
  and zero refractive index acoustic materials},\ }\href
  {https://doi.org/10.1063/1.3686907} {\bibfield  {journal} {\bibinfo
  {journal} {Applied Physics Letters}\ }\textbf {\bibinfo {volume} {100}},\
  \bibinfo {pages} {071911} (\bibinfo {year} {2012})}\BibitemShut {NoStop}%
\bibitem [{\citenamefont {Li}\ \emph {et~al.}(2014)\citenamefont {Li},
  \citenamefont {Wu},\ and\ \citenamefont {Mei}}]{LiApplPhysLett2014}%
  \BibitemOpen
  \bibfield  {author} {\bibinfo {author} {\bibfnamefont {Y.}~\bibnamefont
  {Li}}, \bibinfo {author} {\bibfnamefont {Y.}~\bibnamefont {Wu}},\ and\
  \bibinfo {author} {\bibfnamefont {J.}~\bibnamefont {Mei}},\ }\bibfield
  {title} {\bibinfo {title} {Double {{Dirac}} cones in phononic crystals},\
  }\href {https://doi.org/10.1063/1.4890304} {\bibfield  {journal} {\bibinfo
  {journal} {Applied Physics Letters}\ }\textbf {\bibinfo {volume} {105}},\
  \bibinfo {pages} {014107} (\bibinfo {year} {2014})}\BibitemShut {NoStop}%
\bibitem [{\citenamefont {Dubois}\ \emph {et~al.}(2017)\citenamefont {Dubois},
  \citenamefont {Shi}, \citenamefont {Zhu}, \citenamefont {Wang},\ and\
  \citenamefont {Zhang}}]{DuboisNatCommun2017}%
  \BibitemOpen
  \bibfield  {author} {\bibinfo {author} {\bibfnamefont {M.}~\bibnamefont
  {Dubois}}, \bibinfo {author} {\bibfnamefont {C.}~\bibnamefont {Shi}},
  \bibinfo {author} {\bibfnamefont {X.}~\bibnamefont {Zhu}}, \bibinfo {author}
  {\bibfnamefont {Y.}~\bibnamefont {Wang}},\ and\ \bibinfo {author}
  {\bibfnamefont {X.}~\bibnamefont {Zhang}},\ }\bibfield  {title} {\bibinfo
  {title} {Observation of acoustic {{Dirac}}-like cone and double zero
  refractive index},\ }\href {https://doi.org/10.1038/ncomms14871} {\bibfield
  {journal} {\bibinfo  {journal} {Nature Communications}\ }\textbf {\bibinfo
  {volume} {8}},\ \bibinfo {pages} {14871} (\bibinfo {year}
  {2017})}\BibitemShut {NoStop}%
\bibitem [{\citenamefont {Liberal}\ and\ \citenamefont
  {Engheta}(2017)}]{LiberalNatPhotonics2017}%
  \BibitemOpen
  \bibfield  {author} {\bibinfo {author} {\bibfnamefont {I.}~\bibnamefont
  {Liberal}}\ and\ \bibinfo {author} {\bibfnamefont {N.}~\bibnamefont
  {Engheta}},\ }\bibfield  {title} {\bibinfo {title} {Near-zero refractive
  index photonics},\ }\href {https://doi.org/10.1038/nphoton.2017.13}
  {\bibfield  {journal} {\bibinfo  {journal} {Nature Photonics}\ }\textbf
  {\bibinfo {volume} {11}},\ \bibinfo {pages} {149} (\bibinfo {year}
  {2017})}\BibitemShut {NoStop}%
\bibitem [{\citenamefont {He}\ \emph {et~al.}(2018)\citenamefont {He},
  \citenamefont {Qiu}, \citenamefont {Ye}, \citenamefont {Cai}, \citenamefont
  {Fan}, \citenamefont {Ke}, \citenamefont {Zhang},\ and\ \citenamefont
  {Liu}}]{HeNature2018}%
  \BibitemOpen
  \bibfield  {author} {\bibinfo {author} {\bibfnamefont {H.}~\bibnamefont
  {He}}, \bibinfo {author} {\bibfnamefont {C.}~\bibnamefont {Qiu}}, \bibinfo
  {author} {\bibfnamefont {L.}~\bibnamefont {Ye}}, \bibinfo {author}
  {\bibfnamefont {X.}~\bibnamefont {Cai}}, \bibinfo {author} {\bibfnamefont
  {X.}~\bibnamefont {Fan}}, \bibinfo {author} {\bibfnamefont {M.}~\bibnamefont
  {Ke}}, \bibinfo {author} {\bibfnamefont {F.}~\bibnamefont {Zhang}},\ and\
  \bibinfo {author} {\bibfnamefont {Z.}~\bibnamefont {Liu}},\ }\bibfield
  {title} {\bibinfo {title} {Topological negative refraction of surface
  acoustic waves in a {{Weyl}} phononic crystal},\ }\href
  {https://doi.org/10.1038/s41586-018-0367-9} {\bibfield  {journal} {\bibinfo
  {journal} {Nature}\ }\textbf {\bibinfo {volume} {560}},\ \bibinfo {pages}
  {61} (\bibinfo {year} {2018})}\BibitemShut {NoStop}%
\bibitem [{\citenamefont {Hao}\ \emph {et~al.}(2010)\citenamefont {Hao},
  \citenamefont {Yan},\ and\ \citenamefont {Qiu}}]{HaoApplPhysLett2010}%
  \BibitemOpen
  \bibfield  {author} {\bibinfo {author} {\bibfnamefont {J.}~\bibnamefont
  {Hao}}, \bibinfo {author} {\bibfnamefont {W.}~\bibnamefont {Yan}},\ and\
  \bibinfo {author} {\bibfnamefont {M.}~\bibnamefont {Qiu}},\ }\bibfield
  {title} {\bibinfo {title} {Super-reflection and cloaking based on zero index
  metamaterial},\ }\href {https://doi.org/10.1063/1.3359428} {\bibfield
  {journal} {\bibinfo  {journal} {Applied Physics Letters}\ }\textbf {\bibinfo
  {volume} {96}},\ \bibinfo {pages} {101109} (\bibinfo {year}
  {2010})}\BibitemShut {NoStop}%
\bibitem [{\citenamefont {Huang}\ \emph {et~al.}(2011)\citenamefont {Huang},
  \citenamefont {Lai}, \citenamefont {Hang}, \citenamefont {Zheng},\ and\
  \citenamefont {Chan}}]{HuangNatMater2011}%
  \BibitemOpen
  \bibfield  {author} {\bibinfo {author} {\bibfnamefont {X.}~\bibnamefont
  {Huang}}, \bibinfo {author} {\bibfnamefont {Y.}~\bibnamefont {Lai}}, \bibinfo
  {author} {\bibfnamefont {Z.~H.}\ \bibnamefont {Hang}}, \bibinfo {author}
  {\bibfnamefont {H.}~\bibnamefont {Zheng}},\ and\ \bibinfo {author}
  {\bibfnamefont {C.~T.}\ \bibnamefont {Chan}},\ }\bibfield  {title} {\bibinfo
  {title} {Dirac cones induced by accidental degeneracy in photonic crystals
  and zero-refractive-index materials},\ }\href
  {https://doi.org/10.1038/nmat3030} {\bibfield  {journal} {\bibinfo  {journal}
  {Nature Materials}\ }\textbf {\bibinfo {volume} {10}},\ \bibinfo {pages}
  {582} (\bibinfo {year} {2011})}\BibitemShut {NoStop}%
\bibitem [{\citenamefont {Mousavi}\ \emph {et~al.}(2015)\citenamefont
  {Mousavi}, \citenamefont {Khanikaev},\ and\ \citenamefont
  {Wang}}]{MousaviNatCommun2015}%
  \BibitemOpen
  \bibfield  {author} {\bibinfo {author} {\bibfnamefont {S.~H.}\ \bibnamefont
  {Mousavi}}, \bibinfo {author} {\bibfnamefont {A.~B.}\ \bibnamefont
  {Khanikaev}},\ and\ \bibinfo {author} {\bibfnamefont {Z.}~\bibnamefont
  {Wang}},\ }\bibfield  {title} {\bibinfo {title} {Topologically protected
  elastic waves in phononic metamaterials},\ }\href
  {https://doi.org/10.1038/ncomms9682} {\bibfield  {journal} {\bibinfo
  {journal} {Nature Communications}\ }\textbf {\bibinfo {volume} {6}},\
  \bibinfo {pages} {8682} (\bibinfo {year} {2015})}\BibitemShut {NoStop}%
\bibitem [{\citenamefont {He}\ \emph {et~al.}(2016)\citenamefont {He},
  \citenamefont {Ni}, \citenamefont {Ge}, \citenamefont {Sun}, \citenamefont
  {Chen}, \citenamefont {Lu}, \citenamefont {Liu},\ and\ \citenamefont
  {Chen}}]{HeNatPhys2016}%
  \BibitemOpen
  \bibfield  {author} {\bibinfo {author} {\bibfnamefont {C.}~\bibnamefont
  {He}}, \bibinfo {author} {\bibfnamefont {X.}~\bibnamefont {Ni}}, \bibinfo
  {author} {\bibfnamefont {H.}~\bibnamefont {Ge}}, \bibinfo {author}
  {\bibfnamefont {X.-C.}\ \bibnamefont {Sun}}, \bibinfo {author} {\bibfnamefont
  {Y.-B.}\ \bibnamefont {Chen}}, \bibinfo {author} {\bibfnamefont {M.-H.}\
  \bibnamefont {Lu}}, \bibinfo {author} {\bibfnamefont {X.-P.}\ \bibnamefont
  {Liu}},\ and\ \bibinfo {author} {\bibfnamefont {Y.-F.}\ \bibnamefont
  {Chen}},\ }\bibfield  {title} {\bibinfo {title} {Acoustic topological
  insulator and robust one-way sound transport},\ }\href
  {https://doi.org/10.1038/nphys3867} {\bibfield  {journal} {\bibinfo
  {journal} {Nature Physics}\ }\textbf {\bibinfo {volume} {12}},\ \bibinfo
  {pages} {1124} (\bibinfo {year} {2016})}\BibitemShut {NoStop}%
\bibitem [{\citenamefont {Wei}\ \emph {et~al.}(2017)\citenamefont {Wei},
  \citenamefont {Tian}, \citenamefont {Zuo}, \citenamefont {Cheng},\ and\
  \citenamefont {Liu}}]{WeiPhysRevB2017}%
  \BibitemOpen
  \bibfield  {author} {\bibinfo {author} {\bibfnamefont {Q.}~\bibnamefont
  {Wei}}, \bibinfo {author} {\bibfnamefont {Y.}~\bibnamefont {Tian}}, \bibinfo
  {author} {\bibfnamefont {S.-Y.}\ \bibnamefont {Zuo}}, \bibinfo {author}
  {\bibfnamefont {Y.}~\bibnamefont {Cheng}},\ and\ \bibinfo {author}
  {\bibfnamefont {X.-J.}\ \bibnamefont {Liu}},\ }\bibfield  {title} {\bibinfo
  {title} {Experimental demonstration of topologically protected efficient
  sound propagation in an acoustic waveguide network},\ }\href
  {https://doi.org/10.1103/PhysRevB.95.094305} {\bibfield  {journal} {\bibinfo
  {journal} {Physical Review B}\ }\textbf {\bibinfo {volume} {95}},\ \bibinfo
  {pages} {094305} (\bibinfo {year} {2017})}\BibitemShut {NoStop}%
\bibitem [{\citenamefont {Checkelsky}\ \emph {et~al.}(2012)\citenamefont
  {Checkelsky}, \citenamefont {Ye}, \citenamefont {Onose}, \citenamefont
  {Iwasa},\ and\ \citenamefont {Tokura}}]{CheckelskyNatPhys2012}%
  \BibitemOpen
  \bibfield  {author} {\bibinfo {author} {\bibfnamefont {J.~G.}\ \bibnamefont
  {Checkelsky}}, \bibinfo {author} {\bibfnamefont {J.}~\bibnamefont {Ye}},
  \bibinfo {author} {\bibfnamefont {Y.}~\bibnamefont {Onose}}, \bibinfo
  {author} {\bibfnamefont {Y.}~\bibnamefont {Iwasa}},\ and\ \bibinfo {author}
  {\bibfnamefont {Y.}~\bibnamefont {Tokura}},\ }\bibfield  {title} {\bibinfo
  {title} {Dirac-fermion-mediated ferromagnetism in a topological insulator},\
  }\href {https://doi.org/10.1038/nphys2388} {\bibfield  {journal} {\bibinfo
  {journal} {Nature Physics}\ }\textbf {\bibinfo {volume} {8}},\ \bibinfo
  {pages} {729} (\bibinfo {year} {2012})}\BibitemShut {NoStop}%
\bibitem [{\citenamefont {Wray}(2012)}]{WrayNatPhys2012}%
  \BibitemOpen
  \bibfield  {author} {\bibinfo {author} {\bibfnamefont {L.~A.}\ \bibnamefont
  {Wray}},\ }\bibfield  {title} {\bibinfo {title} {Topological transistor},\
  }\href {https://doi.org/10.1038/nphys2410} {\bibfield  {journal} {\bibinfo
  {journal} {Nature Physics}\ }\textbf {\bibinfo {volume} {8}},\ \bibinfo
  {pages} {705} (\bibinfo {year} {2012})}\BibitemShut {NoStop}%
\bibitem [{\citenamefont {Collins}\ \emph {et~al.}(2018)\citenamefont
  {Collins}, \citenamefont {Tadich}, \citenamefont {Wu}, \citenamefont {Gomes},
  \citenamefont {Rodrigues}, \citenamefont {Liu}, \citenamefont {Hellerstedt},
  \citenamefont {Ryu}, \citenamefont {Tang}, \citenamefont {Mo}, \citenamefont
  {Adam}, \citenamefont {Yang}, \citenamefont {Fuhrer},\ and\ \citenamefont
  {Edmonds}}]{CTW18}%
  \BibitemOpen
  \bibfield  {author} {\bibinfo {author} {\bibfnamefont {J.~L.}\ \bibnamefont
  {Collins}}, \bibinfo {author} {\bibfnamefont {A.}~\bibnamefont {Tadich}},
  \bibinfo {author} {\bibfnamefont {W.}~\bibnamefont {Wu}}, \bibinfo {author}
  {\bibfnamefont {L.~C.}\ \bibnamefont {Gomes}}, \bibinfo {author}
  {\bibfnamefont {J.~N.~B.}\ \bibnamefont {Rodrigues}}, \bibinfo {author}
  {\bibfnamefont {C.}~\bibnamefont {Liu}}, \bibinfo {author} {\bibfnamefont
  {J.}~\bibnamefont {Hellerstedt}}, \bibinfo {author} {\bibfnamefont
  {H.}~\bibnamefont {Ryu}}, \bibinfo {author} {\bibfnamefont {S.}~\bibnamefont
  {Tang}}, \bibinfo {author} {\bibfnamefont {S.-K.}\ \bibnamefont {Mo}},
  \bibinfo {author} {\bibfnamefont {S.}~\bibnamefont {Adam}}, \bibinfo {author}
  {\bibfnamefont {S.~A.}\ \bibnamefont {Yang}}, \bibinfo {author}
  {\bibfnamefont {M.~S.}\ \bibnamefont {Fuhrer}},\ and\ \bibinfo {author}
  {\bibfnamefont {M.~T.}\ \bibnamefont {Edmonds}},\ }\bibfield  {title}
  {\bibinfo {title} {Electric-field-tuned topological phase transition in
  ultrathin {{Na$_3$Bi}}},\ }\href {https://doi.org/10.1038/s41586-018-0788-5}
  {\bibfield  {journal} {\bibinfo  {journal} {Nature}\ }\textbf {\bibinfo
  {volume} {564}},\ \bibinfo {pages} {390} (\bibinfo {year}
  {2018})}\BibitemShut {NoStop}%
\bibitem [{\citenamefont {Wang}\ \emph {et~al.}(2009)\citenamefont {Wang},
  \citenamefont {Chong}, \citenamefont {Joannopoulos},\ and\ \citenamefont
  {Solja{\v c}i\'c}}]{WangNature2009}%
  \BibitemOpen
  \bibfield  {author} {\bibinfo {author} {\bibfnamefont {Z.}~\bibnamefont
  {Wang}}, \bibinfo {author} {\bibfnamefont {Y.}~\bibnamefont {Chong}},
  \bibinfo {author} {\bibfnamefont {J.~D.}\ \bibnamefont {Joannopoulos}},\ and\
  \bibinfo {author} {\bibfnamefont {M.}~\bibnamefont {Solja{\v c}i\'c}},\
  }\bibfield  {title} {\bibinfo {title} {Observation of unidirectional
  backscattering-immune topological electromagnetic states},\ }\href
  {https://doi.org/10.1038/nature08293} {\bibfield  {journal} {\bibinfo
  {journal} {Nature}\ }\textbf {\bibinfo {volume} {461}},\ \bibinfo {pages}
  {772} (\bibinfo {year} {2009})}\BibitemShut {NoStop}%
\bibitem [{\citenamefont {Fleury}\ \emph {et~al.}(2014)\citenamefont {Fleury},
  \citenamefont {Sounas}, \citenamefont {Sieck}, \citenamefont {Haberman},\
  and\ \citenamefont {Alu}}]{FleuryScience2014}%
  \BibitemOpen
  \bibfield  {author} {\bibinfo {author} {\bibfnamefont {R.}~\bibnamefont
  {Fleury}}, \bibinfo {author} {\bibfnamefont {D.~L.}\ \bibnamefont {Sounas}},
  \bibinfo {author} {\bibfnamefont {C.~F.}\ \bibnamefont {Sieck}}, \bibinfo
  {author} {\bibfnamefont {M.~R.}\ \bibnamefont {Haberman}},\ and\ \bibinfo
  {author} {\bibfnamefont {A.}~\bibnamefont {Alu}},\ }\bibfield  {title}
  {\bibinfo {title} {Sound {{Isolation}} and {{Giant Linear Nonreciprocity}} in
  a {{Compact Acoustic Circulator}}},\ }\href
  {https://doi.org/10.1126/science.1246957} {\bibfield  {journal} {\bibinfo
  {journal} {Science}\ }\textbf {\bibinfo {volume} {343}},\ \bibinfo {pages}
  {516} (\bibinfo {year} {2014})}\BibitemShut {NoStop}%
\bibitem [{\citenamefont {Yang}\ \emph {et~al.}(2015)\citenamefont {Yang},
  \citenamefont {Gao}, \citenamefont {Shi}, \citenamefont {Lin}, \citenamefont
  {Gao}, \citenamefont {Chong},\ and\ \citenamefont
  {Zhang}}]{YangPhysRevLett2015}%
  \BibitemOpen
  \bibfield  {author} {\bibinfo {author} {\bibfnamefont {Z.}~\bibnamefont
  {Yang}}, \bibinfo {author} {\bibfnamefont {F.}~\bibnamefont {Gao}}, \bibinfo
  {author} {\bibfnamefont {X.}~\bibnamefont {Shi}}, \bibinfo {author}
  {\bibfnamefont {X.}~\bibnamefont {Lin}}, \bibinfo {author} {\bibfnamefont
  {Z.}~\bibnamefont {Gao}}, \bibinfo {author} {\bibfnamefont {Y.}~\bibnamefont
  {Chong}},\ and\ \bibinfo {author} {\bibfnamefont {B.}~\bibnamefont {Zhang}},\
  }\bibfield  {title} {\bibinfo {title} {Topological {{Acoustics}}},\ }\href
  {https://doi.org/10.1103/PhysRevLett.114.114301} {\bibfield  {journal}
  {\bibinfo  {journal} {Physical Review Letters}\ }\textbf {\bibinfo {volume}
  {114}},\ \bibinfo {pages} {114301} (\bibinfo {year} {2015})}\BibitemShut
  {NoStop}%
\bibitem [{\citenamefont {Wang}\ \emph {et~al.}(2015)\citenamefont {Wang},
  \citenamefont {Lu},\ and\ \citenamefont {Bertoldi}}]{WangPhysRevLett2015}%
  \BibitemOpen
  \bibfield  {author} {\bibinfo {author} {\bibfnamefont {P.}~\bibnamefont
  {Wang}}, \bibinfo {author} {\bibfnamefont {L.}~\bibnamefont {Lu}},\ and\
  \bibinfo {author} {\bibfnamefont {K.}~\bibnamefont {Bertoldi}},\ }\bibfield
  {title} {\bibinfo {title} {Topological {{Phononic Crystals}} with
  {{One}}-{{Way Elastic Edge Waves}}},\ }\href
  {https://doi.org/10.1103/PhysRevLett.115.104302} {\bibfield  {journal}
  {\bibinfo  {journal} {Physical Review Letters}\ }\textbf {\bibinfo {volume}
  {115}},\ \bibinfo {pages} {104302} (\bibinfo {year} {2015})}\BibitemShut
  {NoStop}%
\bibitem [{\citenamefont {Nash}\ \emph {et~al.}(2015)\citenamefont {Nash},
  \citenamefont {Kleckner}, \citenamefont {Read}, \citenamefont {Vitelli},
  \citenamefont {Turner},\ and\ \citenamefont
  {Irvine}}]{NashProcNatlAcadSci2015}%
  \BibitemOpen
  \bibfield  {author} {\bibinfo {author} {\bibfnamefont {L.~M.}\ \bibnamefont
  {Nash}}, \bibinfo {author} {\bibfnamefont {D.}~\bibnamefont {Kleckner}},
  \bibinfo {author} {\bibfnamefont {A.}~\bibnamefont {Read}}, \bibinfo {author}
  {\bibfnamefont {V.}~\bibnamefont {Vitelli}}, \bibinfo {author} {\bibfnamefont
  {A.~M.}\ \bibnamefont {Turner}},\ and\ \bibinfo {author} {\bibfnamefont
  {W.~T.~M.}\ \bibnamefont {Irvine}},\ }\bibfield  {title} {\bibinfo {title}
  {Topological mechanics of gyroscopic metamaterials},\ }\href
  {https://doi.org/10.1073/pnas.1507413112} {\bibfield  {journal} {\bibinfo
  {journal} {Proceedings of the National Academy of Sciences}\ }\textbf
  {\bibinfo {volume} {112}},\ \bibinfo {pages} {14495} (\bibinfo {year}
  {2015})}\BibitemShut {NoStop}%
\bibitem [{\citenamefont {Wu}\ and\ \citenamefont
  {Hu}(2015)}]{WuPhysRevLett2015}%
  \BibitemOpen
  \bibfield  {author} {\bibinfo {author} {\bibfnamefont {L.-H.}\ \bibnamefont
  {Wu}}\ and\ \bibinfo {author} {\bibfnamefont {X.}~\bibnamefont {Hu}},\
  }\bibfield  {title} {\bibinfo {title} {Scheme for {{Achieving}} a
  {{Topological Photonic Crystal}} by {{Using Dielectric Material}}},\ }\href
  {https://doi.org/10.1103/PhysRevLett.114.223901} {\bibfield  {journal}
  {\bibinfo  {journal} {Physical Review Letters}\ }\textbf {\bibinfo {volume}
  {114}},\ \bibinfo {pages} {223901} (\bibinfo {year} {2015})}\BibitemShut
  {NoStop}%
\bibitem [{\citenamefont {Zhang}\ \emph
  {et~al.}(2017{\natexlab{a}})\citenamefont {Zhang}, \citenamefont {Wei},
  \citenamefont {Cheng}, \citenamefont {Zhang}, \citenamefont {Wu},\ and\
  \citenamefont {Liu}}]{ZhangPhysRevLett2017}%
  \BibitemOpen
  \bibfield  {author} {\bibinfo {author} {\bibfnamefont {Z.}~\bibnamefont
  {Zhang}}, \bibinfo {author} {\bibfnamefont {Q.}~\bibnamefont {Wei}}, \bibinfo
  {author} {\bibfnamefont {Y.}~\bibnamefont {Cheng}}, \bibinfo {author}
  {\bibfnamefont {T.}~\bibnamefont {Zhang}}, \bibinfo {author} {\bibfnamefont
  {D.}~\bibnamefont {Wu}},\ and\ \bibinfo {author} {\bibfnamefont
  {X.}~\bibnamefont {Liu}},\ }\bibfield  {title} {\bibinfo {title} {Topological
  {{Creation}} of {{Acoustic Pseudospin Multipoles}} in a {{Flow}}-{{Free
  Symmetry}}-{{Broken Metamaterial Lattice}}},\ }\href
  {https://doi.org/10.1103/PhysRevLett.118.084303} {\bibfield  {journal}
  {\bibinfo  {journal} {Physical Review Letters}\ }\textbf {\bibinfo {volume}
  {118}},\ \bibinfo {pages} {084303} (\bibinfo {year}
  {2017}{\natexlab{a}})}\BibitemShut {NoStop}%
\bibitem [{\citenamefont {Zhang}\ \emph
  {et~al.}(2017{\natexlab{b}})\citenamefont {Zhang}, \citenamefont {Tian},
  \citenamefont {Cheng}, \citenamefont {Liu},\ and\ \citenamefont
  {Christensen}}]{ZhangPhysRevB2017}%
  \BibitemOpen
  \bibfield  {author} {\bibinfo {author} {\bibfnamefont {Z.}~\bibnamefont
  {Zhang}}, \bibinfo {author} {\bibfnamefont {Y.}~\bibnamefont {Tian}},
  \bibinfo {author} {\bibfnamefont {Y.}~\bibnamefont {Cheng}}, \bibinfo
  {author} {\bibfnamefont {X.}~\bibnamefont {Liu}},\ and\ \bibinfo {author}
  {\bibfnamefont {J.}~\bibnamefont {Christensen}},\ }\bibfield  {title}
  {\bibinfo {title} {Experimental verification of acoustic pseudospin
  multipoles in a symmetry-broken snowflakelike topological insulator},\ }\href
  {https://doi.org/10.1103/PhysRevB.96.241306} {\bibfield  {journal} {\bibinfo
  {journal} {Physical Review B}\ }\textbf {\bibinfo {volume} {96}},\ \bibinfo
  {pages} {241306(R)} (\bibinfo {year} {2017}{\natexlab{b}})}\BibitemShut
  {NoStop}%
\bibitem [{\citenamefont {Deng}\ \emph {et~al.}(2017)\citenamefont {Deng},
  \citenamefont {Ge}, \citenamefont {Tian}, \citenamefont {Lu},\ and\
  \citenamefont {Jing}}]{DengPhysRevB2017}%
  \BibitemOpen
  \bibfield  {author} {\bibinfo {author} {\bibfnamefont {Y.}~\bibnamefont
  {Deng}}, \bibinfo {author} {\bibfnamefont {H.}~\bibnamefont {Ge}}, \bibinfo
  {author} {\bibfnamefont {Y.}~\bibnamefont {Tian}}, \bibinfo {author}
  {\bibfnamefont {M.}~\bibnamefont {Lu}},\ and\ \bibinfo {author}
  {\bibfnamefont {Y.}~\bibnamefont {Jing}},\ }\bibfield  {title} {\bibinfo
  {title} {Observation of zone folding induced acoustic topological insulators
  and the role of spin-mixing defects},\ }\href
  {https://doi.org/10.1103/PhysRevB.96.184305} {\bibfield  {journal} {\bibinfo
  {journal} {Physical Review B}\ }\textbf {\bibinfo {volume} {96}},\ \bibinfo
  {pages} {184305} (\bibinfo {year} {2017})}\BibitemShut {NoStop}%
\bibitem [{\citenamefont {Yang}\ \emph {et~al.}(2018)\citenamefont {Yang},
  \citenamefont {Xu}, \citenamefont {Xu}, \citenamefont {Wang}, \citenamefont
  {Jiang}, \citenamefont {Hu},\ and\ \citenamefont
  {Hang}}]{YangPhysRevLett2018}%
  \BibitemOpen
  \bibfield  {author} {\bibinfo {author} {\bibfnamefont {Y.}~\bibnamefont
  {Yang}}, \bibinfo {author} {\bibfnamefont {Y.~F.}\ \bibnamefont {Xu}},
  \bibinfo {author} {\bibfnamefont {T.}~\bibnamefont {Xu}}, \bibinfo {author}
  {\bibfnamefont {H.-X.}\ \bibnamefont {Wang}}, \bibinfo {author}
  {\bibfnamefont {J.-H.}\ \bibnamefont {Jiang}}, \bibinfo {author}
  {\bibfnamefont {X.}~\bibnamefont {Hu}},\ and\ \bibinfo {author}
  {\bibfnamefont {Z.~H.}\ \bibnamefont {Hang}},\ }\bibfield  {title} {\bibinfo
  {title} {Visualization of a {{Unidirectional Electromagnetic Waveguide Using
  Topological Photonic Crystals Made}} of {{Dielectric Materials}}},\ }\href
  {https://doi.org/10.1103/PhysRevLett.120.217401} {\bibfield  {journal}
  {\bibinfo  {journal} {Physical Review Letters}\ }\textbf {\bibinfo {volume}
  {120}},\ \bibinfo {pages} {217401} (\bibinfo {year} {2018})}\BibitemShut
  {NoStop}%
\bibitem [{\citenamefont {Xia}\ \emph {et~al.}(2017)\citenamefont {Xia},
  \citenamefont {Liu}, \citenamefont {Huang}, \citenamefont {Dai},
  \citenamefont {Jiao}, \citenamefont {Zang}, \citenamefont {Yu}, \citenamefont
  {Zheng},\ and\ \citenamefont {Liu}}]{XiaPhysRevB2017}%
  \BibitemOpen
  \bibfield  {author} {\bibinfo {author} {\bibfnamefont {B.-Z.}\ \bibnamefont
  {Xia}}, \bibinfo {author} {\bibfnamefont {T.-T.}\ \bibnamefont {Liu}},
  \bibinfo {author} {\bibfnamefont {G.-L.}\ \bibnamefont {Huang}}, \bibinfo
  {author} {\bibfnamefont {H.-Q.}\ \bibnamefont {Dai}}, \bibinfo {author}
  {\bibfnamefont {J.-R.}\ \bibnamefont {Jiao}}, \bibinfo {author}
  {\bibfnamefont {X.-G.}\ \bibnamefont {Zang}}, \bibinfo {author}
  {\bibfnamefont {D.-J.}\ \bibnamefont {Yu}}, \bibinfo {author} {\bibfnamefont
  {S.-J.}\ \bibnamefont {Zheng}},\ and\ \bibinfo {author} {\bibfnamefont
  {J.}~\bibnamefont {Liu}},\ }\bibfield  {title} {\bibinfo {title} {Topological
  phononic insulator with robust pseudospin-dependent transport},\ }\href
  {https://doi.org/10.1103/PhysRevB.96.094106} {\bibfield  {journal} {\bibinfo
  {journal} {Physical Review B}\ }\textbf {\bibinfo {volume} {96}},\ \bibinfo
  {pages} {094106} (\bibinfo {year} {2017})}\BibitemShut {NoStop}%
\bibitem [{\citenamefont {Mei}\ \emph {et~al.}(2012)\citenamefont {Mei},
  \citenamefont {Wu}, \citenamefont {Chan},\ and\ \citenamefont
  {Zhang}}]{MeiPhysRevB2012}%
  \BibitemOpen
  \bibfield  {author} {\bibinfo {author} {\bibfnamefont {J.}~\bibnamefont
  {Mei}}, \bibinfo {author} {\bibfnamefont {Y.}~\bibnamefont {Wu}}, \bibinfo
  {author} {\bibfnamefont {C.~T.}\ \bibnamefont {Chan}},\ and\ \bibinfo
  {author} {\bibfnamefont {Z.-Q.}\ \bibnamefont {Zhang}},\ }\bibfield  {title}
  {\bibinfo {title} {First-principles study of {{Dirac}} and {{Dirac}}-like
  cones in phononic and photonic crystals},\ }\href
  {https://doi.org/10.1103/PhysRevB.86.035141} {\bibfield  {journal} {\bibinfo
  {journal} {Physical Review B}\ }\textbf {\bibinfo {volume} {86}},\ \bibinfo
  {pages} {035141} (\bibinfo {year} {2012})}\BibitemShut {NoStop}%
\bibitem [{\citenamefont {Sakoda}(2012)}]{SakodaOptExpress2012}%
  \BibitemOpen
  \bibfield  {author} {\bibinfo {author} {\bibfnamefont {K.}~\bibnamefont
  {Sakoda}},\ }\bibfield  {title} {\bibinfo {title} {Double {{Dirac}} cones in
  triangular-lattice metamaterials},\ }\href
  {https://doi.org/10.1364/OE.20.009925} {\bibfield  {journal} {\bibinfo
  {journal} {Optics Express}\ }\textbf {\bibinfo {volume} {20}},\ \bibinfo
  {pages} {9925} (\bibinfo {year} {2012})}\BibitemShut {NoStop}%
\bibitem [{\citenamefont {Chen}\ \emph {et~al.}(2015)\citenamefont {Chen},
  \citenamefont {Ni}, \citenamefont {Wu}, \citenamefont {He}, \citenamefont
  {Sun}, \citenamefont {Zheng}, \citenamefont {Lu},\ and\ \citenamefont
  {Chen}}]{ChenSciRep2015}%
  \BibitemOpen
  \bibfield  {author} {\bibinfo {author} {\bibfnamefont {Z.-G.}\ \bibnamefont
  {Chen}}, \bibinfo {author} {\bibfnamefont {X.}~\bibnamefont {Ni}}, \bibinfo
  {author} {\bibfnamefont {Y.}~\bibnamefont {Wu}}, \bibinfo {author}
  {\bibfnamefont {C.}~\bibnamefont {He}}, \bibinfo {author} {\bibfnamefont
  {X.-C.}\ \bibnamefont {Sun}}, \bibinfo {author} {\bibfnamefont {L.-Y.}\
  \bibnamefont {Zheng}}, \bibinfo {author} {\bibfnamefont {M.-H.}\ \bibnamefont
  {Lu}},\ and\ \bibinfo {author} {\bibfnamefont {Y.-F.}\ \bibnamefont {Chen}},\
  }\bibfield  {title} {\bibinfo {title} {Accidental degeneracy of double
  {{Dirac}} cones in a phononic crystal},\ }\href
  {https://doi.org/10.1038/srep04613} {\bibfield  {journal} {\bibinfo
  {journal} {Scientific Reports}\ }\textbf {\bibinfo {volume} {4}},\ \bibinfo
  {pages} {4613} (\bibinfo {year} {2015})}\BibitemShut {NoStop}%
\bibitem [{\citenamefont {Mei}\ \emph {et~al.}(2016)\citenamefont {Mei},
  \citenamefont {Chen},\ and\ \citenamefont {Wu}}]{MeiSciRep2016}%
  \BibitemOpen
  \bibfield  {author} {\bibinfo {author} {\bibfnamefont {J.}~\bibnamefont
  {Mei}}, \bibinfo {author} {\bibfnamefont {Z.}~\bibnamefont {Chen}},\ and\
  \bibinfo {author} {\bibfnamefont {Y.}~\bibnamefont {Wu}},\ }\bibfield
  {title} {\bibinfo {title} {Pseudo-time-reversal symmetry and topological edge
  states in two-dimensional acoustic crystals},\ }\href
  {https://doi.org/10.1038/srep32752} {\bibfield  {journal} {\bibinfo
  {journal} {Scientific Reports}\ }\textbf {\bibinfo {volume} {6}},\ \bibinfo
  {pages} {32752} (\bibinfo {year} {2016})}\BibitemShut {NoStop}%
\bibitem [{\citenamefont {S{\"u}sstrunk}\ \emph {et~al.}(2017)\citenamefont
  {S{\"u}sstrunk}, \citenamefont {Zimmermann},\ and\ \citenamefont
  {Huber}}]{SZH17}%
  \BibitemOpen
  \bibfield  {author} {\bibinfo {author} {\bibfnamefont {R.}~\bibnamefont
  {S{\"u}sstrunk}}, \bibinfo {author} {\bibfnamefont {P.}~\bibnamefont
  {Zimmermann}},\ and\ \bibinfo {author} {\bibfnamefont {S.~D.}\ \bibnamefont
  {Huber}},\ }\bibfield  {title} {\bibinfo {title} {Switchable topological
  phonon channels},\ }\href {https://doi.org/10.1088/1367-2630/aa591c}
  {\bibfield  {journal} {\bibinfo  {journal} {New Journal of Physics}\ }\textbf
  {\bibinfo {volume} {19}},\ \bibinfo {pages} {015013} (\bibinfo {year}
  {2017})}\BibitemShut {NoStop}%
\bibitem [{\citenamefont {Zhang}\ \emph {et~al.}(2018)\citenamefont {Zhang},
  \citenamefont {Tian}, \citenamefont {Cheng}, \citenamefont {Wei},
  \citenamefont {Liu},\ and\ \citenamefont
  {Christensen}}]{ZhangPhysRevAppl2018}%
  \BibitemOpen
  \bibfield  {author} {\bibinfo {author} {\bibfnamefont {Z.}~\bibnamefont
  {Zhang}}, \bibinfo {author} {\bibfnamefont {Y.}~\bibnamefont {Tian}},
  \bibinfo {author} {\bibfnamefont {Y.}~\bibnamefont {Cheng}}, \bibinfo
  {author} {\bibfnamefont {Q.}~\bibnamefont {Wei}}, \bibinfo {author}
  {\bibfnamefont {X.}~\bibnamefont {Liu}},\ and\ \bibinfo {author}
  {\bibfnamefont {J.}~\bibnamefont {Christensen}},\ }\bibfield  {title}
  {\bibinfo {title} {Topological {{Acoustic Delay Line}}},\ }\href
  {https://doi.org/10.1103/PhysRevApplied.9.034032} {\bibfield  {journal}
  {\bibinfo  {journal} {Physical Review Applied}\ }\textbf {\bibinfo {volume}
  {9}},\ \bibinfo {pages} {034032} (\bibinfo {year} {2018})}\BibitemShut
  {NoStop}%
\bibitem [{\citenamefont {Xia}\ \emph {et~al.}(2018)\citenamefont {Xia},
  \citenamefont {Jia}, \citenamefont {Sun}, \citenamefont {Yuan}, \citenamefont
  {Ge}, \citenamefont {Si},\ and\ \citenamefont {Liu}}]{XJS18}%
  \BibitemOpen
  \bibfield  {author} {\bibinfo {author} {\bibfnamefont {J.-P.}\ \bibnamefont
  {Xia}}, \bibinfo {author} {\bibfnamefont {D.}~\bibnamefont {Jia}}, \bibinfo
  {author} {\bibfnamefont {H.-X.}\ \bibnamefont {Sun}}, \bibinfo {author}
  {\bibfnamefont {S.-Q.}\ \bibnamefont {Yuan}}, \bibinfo {author}
  {\bibfnamefont {Y.}~\bibnamefont {Ge}}, \bibinfo {author} {\bibfnamefont
  {Q.-R.}\ \bibnamefont {Si}},\ and\ \bibinfo {author} {\bibfnamefont {X.-J.}\
  \bibnamefont {Liu}},\ }\bibfield  {title} {\bibinfo {title} {Programmable
  {{Coding Acoustic Topological Insulator}}},\ }\href
  {https://doi.org/10.1002/adma.201805002} {\bibfield  {journal} {\bibinfo
  {journal} {Advanced Materials}\ }\textbf {\bibinfo {volume} {30}},\ \bibinfo
  {pages} {1805002} (\bibinfo {year} {2018})}\BibitemShut {NoStop}%
\bibitem [{\citenamefont {Boatti}\ \emph {et~al.}(2017)\citenamefont {Boatti},
  \citenamefont {Vasios},\ and\ \citenamefont {Bertoldi}}]{BoattiAdvMater2017}%
  \BibitemOpen
  \bibfield  {author} {\bibinfo {author} {\bibfnamefont {E.}~\bibnamefont
  {Boatti}}, \bibinfo {author} {\bibfnamefont {N.}~\bibnamefont {Vasios}},\
  and\ \bibinfo {author} {\bibfnamefont {K.}~\bibnamefont {Bertoldi}},\
  }\bibfield  {title} {\bibinfo {title} {Origami {{Metamaterials}} for
  {{Tunable Thermal Expansion}}},\ }\href
  {https://doi.org/10.1002/adma.201700360} {\bibfield  {journal} {\bibinfo
  {journal} {Advanced Materials}\ }\textbf {\bibinfo {volume} {29}},\ \bibinfo
  {pages} {1700360} (\bibinfo {year} {2017})}\BibitemShut {NoStop}%
\bibitem [{\citenamefont {Oliveira}\ \emph {et~al.}(2016)\citenamefont
  {Oliveira}, \citenamefont {Silva}, \citenamefont {Morais}, \citenamefont
  {Alvarenga},\ and\ \citenamefont {F\'elix}}]{OliveiraJPhysConfSer2016}%
  \BibitemOpen
  \bibfield  {author} {\bibinfo {author} {\bibfnamefont {P.~A.}\ \bibnamefont
  {Oliveira}}, \bibinfo {author} {\bibfnamefont {R.~M.~B.}\ \bibnamefont
  {Silva}}, \bibinfo {author} {\bibfnamefont {G.~C.}\ \bibnamefont {Morais}},
  \bibinfo {author} {\bibfnamefont {A.~V.}\ \bibnamefont {Alvarenga}},\ and\
  \bibinfo {author} {\bibfnamefont {R.~P. B.~C.}\ \bibnamefont {F\'elix}},\
  }\bibfield  {title} {\bibinfo {title} {Speed of sound as a function of
  temperature for ultrasonic propagation in soybean oil},\ }\href
  {https://doi.org/10.1088/1742-6596/733/1/012040} {\bibfield  {journal}
  {\bibinfo  {journal} {Journal of Physics: Conference Series}\ }\textbf
  {\bibinfo {volume} {733}},\ \bibinfo {pages} {012040} (\bibinfo {year}
  {2016})}\BibitemShut {NoStop}%
\bibitem [{\citenamefont {Yu}\ \emph {et~al.}(2016)\citenamefont {Yu},
  \citenamefont {Sun}, \citenamefont {Ni}, \citenamefont {Wang}, \citenamefont
  {Yan}, \citenamefont {He}, \citenamefont {Liu}, \citenamefont {Feng},
  \citenamefont {Lu},\ and\ \citenamefont {Chen}}]{YSN16}%
  \BibitemOpen
  \bibfield  {author} {\bibinfo {author} {\bibfnamefont {S.-Y.}\ \bibnamefont
  {Yu}}, \bibinfo {author} {\bibfnamefont {X.-C.}\ \bibnamefont {Sun}},
  \bibinfo {author} {\bibfnamefont {X.}~\bibnamefont {Ni}}, \bibinfo {author}
  {\bibfnamefont {Q.}~\bibnamefont {Wang}}, \bibinfo {author} {\bibfnamefont
  {X.-J.}\ \bibnamefont {Yan}}, \bibinfo {author} {\bibfnamefont
  {C.}~\bibnamefont {He}}, \bibinfo {author} {\bibfnamefont {X.-P.}\
  \bibnamefont {Liu}}, \bibinfo {author} {\bibfnamefont {L.}~\bibnamefont
  {Feng}}, \bibinfo {author} {\bibfnamefont {M.-H.}\ \bibnamefont {Lu}},\ and\
  \bibinfo {author} {\bibfnamefont {Y.-F.}\ \bibnamefont {Chen}},\ }\bibfield
  {title} {\bibinfo {title} {Surface phononic graphene},\ }\href
  {https://doi.org/10.1038/nmat4743} {\bibfield  {journal} {\bibinfo  {journal}
  {Nature Materials}\ }\textbf {\bibinfo {volume} {15}},\ \bibinfo {pages}
  {1243} (\bibinfo {year} {2016})}\BibitemShut {NoStop}%
\bibitem [{\citenamefont {Polini}\ \emph {et~al.}(2013)\citenamefont {Polini},
  \citenamefont {Guinea}, \citenamefont {Lewenstein}, \citenamefont
  {Manoharan},\ and\ \citenamefont {Pellegrini}}]{PoliniNatNano2013}%
  \BibitemOpen
  \bibfield  {author} {\bibinfo {author} {\bibfnamefont {M.}~\bibnamefont
  {Polini}}, \bibinfo {author} {\bibfnamefont {F.}~\bibnamefont {Guinea}},
  \bibinfo {author} {\bibfnamefont {M.}~\bibnamefont {Lewenstein}}, \bibinfo
  {author} {\bibfnamefont {H.~C.}\ \bibnamefont {Manoharan}},\ and\ \bibinfo
  {author} {\bibfnamefont {V.}~\bibnamefont {Pellegrini}},\ }\bibfield  {title}
  {\bibinfo {title} {Artificial honeycomb lattices for electrons, atoms and
  photons},\ }\href {https://doi.org/10.1038/nnano.2013.161} {\bibfield
  {journal} {\bibinfo  {journal} {Nature Nanotechnology}\ }\textbf {\bibinfo
  {volume} {8}},\ \bibinfo {pages} {625} (\bibinfo {year} {2013})}\BibitemShut
  {NoStop}%
\end{thebibliography}%


\UseRawInputEncoding
\begin{thebibliography}{10}%
\makeatletter
\providecommand \@ifxundefined [1]{%
 \@ifx{#1\undefined}
}%
\providecommand \@ifnum [1]{%
 \ifnum #1\expandafter \@firstoftwo
 \else \expandafter \@secondoftwo
 \fi
}%
\providecommand \@ifx [1]{%
 \ifx #1\expandafter \@firstoftwo
 \else \expandafter \@secondoftwo
 \fi
}%
\providecommand \natexlab [1]{#1}%
\providecommand \enquote  [1]{``#1''}%
\providecommand \bibnamefont  [1]{#1}%
\providecommand \bibfnamefont [1]{#1}%
\providecommand \citenamefont [1]{#1}%
\providecommand \href@noop [0]{\@secondoftwo}%
\providecommand \href [0]{\begingroup \@sanitize@url \@href}%
\providecommand \@href[1]{\@@startlink{#1}\@@href}%
\providecommand \@@href[1]{\endgroup#1\@@endlink}%
\providecommand \@sanitize@url [0]{\catcode `\\12\catcode `\$12\catcode
  `\&12\catcode `\#12\catcode `\^12\catcode `\_12\catcode `\%12\relax}%
\providecommand \@@startlink[1]{}%
\providecommand \@@endlink[0]{}%
\providecommand \url  [0]{\begingroup\@sanitize@url \@url }%
\providecommand \@url [1]{\endgroup\@href {#1}{\urlprefix }}%
\providecommand \urlprefix  [0]{URL }%
\providecommand \Eprint [0]{\href }%
\providecommand \doibase [0]{https://doi.org/}%
\providecommand \selectlanguage [0]{\@gobble}%
\providecommand \bibinfo  [0]{\@secondoftwo}%
\providecommand \bibfield  [0]{\@secondoftwo}%
\providecommand \translation [1]{[#1]}%
\providecommand \BibitemOpen [0]{}%
\providecommand \bibitemStop [0]{}%
\providecommand \bibitemNoStop [0]{.\EOS\space}%
\providecommand \EOS [0]{\spacefactor3000\relax}%
\providecommand \BibitemShut  [1]{\csname bibitem#1\endcsname}%
\let\auto@bib@innerbib\@empty
\bibitem [{\citenamefont {Symko}\ \emph {et~al.}(2004)\citenamefont {Symko},
  \citenamefont {{Abdel-Rahman}}, \citenamefont {Kwon}, \citenamefont {Emmi},\
  and\ \citenamefont {Behunin}}]{SAK04}%
  \BibitemOpen
  \bibfield  {author} {\bibinfo {author} {\bibfnamefont {O.}~\bibnamefont
  {Symko}}, \bibinfo {author} {\bibfnamefont {E.}~\bibnamefont
  {{Abdel-Rahman}}}, \bibinfo {author} {\bibfnamefont {Y.}~\bibnamefont
  {Kwon}}, \bibinfo {author} {\bibfnamefont {M.}~\bibnamefont {Emmi}},\ and\
  \bibinfo {author} {\bibfnamefont {R.}~\bibnamefont {Behunin}},\ }\bibfield
  {title} {\bibinfo {title} {Design and development of high-frequency
  thermoacoustic engines for thermal management in microelectronics},\ }\href
  {https://doi.org/10.1016/j.mejo.2003.09.017} {\bibfield  {journal} {\bibinfo
  {journal} {Microelectronics Journal}\ }\textbf {\bibinfo {volume} {35}},\
  \bibinfo {pages} {185} (\bibinfo {year} {2004})}\BibitemShut {NoStop}%
\bibitem [{\citenamefont {Chen}\ \emph {et~al.}(2021)\citenamefont {Chen},
  \citenamefont {Tang}, \citenamefont {Mace},\ and\ \citenamefont
  {Yu}}]{CTM21}%
  \BibitemOpen
  \bibfield  {author} {\bibinfo {author} {\bibfnamefont {G.}~\bibnamefont
  {Chen}}, \bibinfo {author} {\bibfnamefont {L.}~\bibnamefont {Tang}}, \bibinfo
  {author} {\bibfnamefont {B.}~\bibnamefont {Mace}},\ and\ \bibinfo {author}
  {\bibfnamefont {Z.}~\bibnamefont {Yu}},\ }\bibfield  {title} {\bibinfo
  {title} {Multi-physics coupling in thermoacoustic devices: A review},\ }\href
  {https://doi.org/10.1016/j.rser.2021.111170} {\bibfield  {journal} {\bibinfo
  {journal} {Renewable and Sustainable Energy Reviews}\ }\textbf {\bibinfo
  {volume} {146}},\ \bibinfo {pages} {111170} (\bibinfo {year}
  {2021})}\BibitemShut {NoStop}%
\bibitem [{\citenamefont {Boonstra}(1950)}]{B50}%
  \BibitemOpen
  \bibfield  {author} {\bibinfo {author} {\bibfnamefont {B.~B. S.~T.}\
  \bibnamefont {Boonstra}},\ }\bibfield  {title} {\bibinfo {title}
  {Stress-€strain properties of natural rubber under biaxial strain},\ }\href
  {https://doi.org/10.1063/1.1699549} {\bibfield  {journal} {\bibinfo
  {journal} {Journal of Applied Physics}\ }\textbf {\bibinfo {volume} {21}},\
  \bibinfo {pages} {1098} (\bibinfo {year} {1950})}\BibitemShut {NoStop}%
\bibitem [{\citenamefont {Wu}\ and\ \citenamefont
  {Hu}(2015)}]{WuPhysRevLett2015}%
  \BibitemOpen
  \bibfield  {author} {\bibinfo {author} {\bibfnamefont {L.-H.}\ \bibnamefont
  {Wu}}\ and\ \bibinfo {author} {\bibfnamefont {X.}~\bibnamefont {Hu}},\
  }\bibfield  {title} {\bibinfo {title} {Scheme for {{Achieving}} a
  {{Topological Photonic Crystal}} by {{Using Dielectric Material}}},\ }\href
  {https://doi.org/10.1103/PhysRevLett.114.223901} {\bibfield  {journal}
  {\bibinfo  {journal} {Physical Review Letters}\ }\textbf {\bibinfo {volume}
  {114}},\ \bibinfo {pages} {223901} (\bibinfo {year} {2015})}\BibitemShut
  {NoStop}%
\bibitem [{\citenamefont {Mei}\ \emph {et~al.}(2012)\citenamefont {Mei},
  \citenamefont {Wu}, \citenamefont {Chan},\ and\ \citenamefont
  {Zhang}}]{MeiPhysRevB2012}%
  \BibitemOpen
  \bibfield  {author} {\bibinfo {author} {\bibfnamefont {J.}~\bibnamefont
  {Mei}}, \bibinfo {author} {\bibfnamefont {Y.}~\bibnamefont {Wu}}, \bibinfo
  {author} {\bibfnamefont {C.~T.}\ \bibnamefont {Chan}},\ and\ \bibinfo
  {author} {\bibfnamefont {Z.-Q.}\ \bibnamefont {Zhang}},\ }\bibfield  {title}
  {\bibinfo {title} {First-principles study of {{Dirac}} and {{Dirac}}-like
  cones in phononic and photonic crystals},\ }\href
  {https://doi.org/10.1103/PhysRevB.86.035141} {\bibfield  {journal} {\bibinfo
  {journal} {Physical Review B}\ }\textbf {\bibinfo {volume} {86}},\ \bibinfo
  {pages} {035141} (\bibinfo {year} {2012})}\BibitemShut {NoStop}%
\bibitem [{\citenamefont {He}\ \emph {et~al.}(2016)\citenamefont {He},
  \citenamefont {Ni}, \citenamefont {Ge}, \citenamefont {Sun}, \citenamefont
  {Chen}, \citenamefont {Lu}, \citenamefont {Liu},\ and\ \citenamefont
  {Chen}}]{HeNatPhys2016}%
  \BibitemOpen
  \bibfield  {author} {\bibinfo {author} {\bibfnamefont {C.}~\bibnamefont
  {He}}, \bibinfo {author} {\bibfnamefont {X.}~\bibnamefont {Ni}}, \bibinfo
  {author} {\bibfnamefont {H.}~\bibnamefont {Ge}}, \bibinfo {author}
  {\bibfnamefont {X.-C.}\ \bibnamefont {Sun}}, \bibinfo {author} {\bibfnamefont
  {Y.-B.}\ \bibnamefont {Chen}}, \bibinfo {author} {\bibfnamefont {M.-H.}\
  \bibnamefont {Lu}}, \bibinfo {author} {\bibfnamefont {X.-P.}\ \bibnamefont
  {Liu}},\ and\ \bibinfo {author} {\bibfnamefont {Y.-F.}\ \bibnamefont
  {Chen}},\ }\bibfield  {title} {\bibinfo {title} {Acoustic topological
  insulator and robust one-way sound transport},\ }\href
  {https://doi.org/10.1038/nphys3867} {\bibfield  {journal} {\bibinfo
  {journal} {Nature Physics}\ }\textbf {\bibinfo {volume} {12}},\ \bibinfo
  {pages} {1124} (\bibinfo {year} {2016})}\BibitemShut {NoStop}%
\bibitem [{\citenamefont {Zhang}\ \emph
  {et~al.}(2017{\natexlab{a}})\citenamefont {Zhang}, \citenamefont {Wei},
  \citenamefont {Cheng}, \citenamefont {Zhang}, \citenamefont {Wu},\ and\
  \citenamefont {Liu}}]{ZhangPhysRevLett2017}%
  \BibitemOpen
  \bibfield  {author} {\bibinfo {author} {\bibfnamefont {Z.}~\bibnamefont
  {Zhang}}, \bibinfo {author} {\bibfnamefont {Q.}~\bibnamefont {Wei}}, \bibinfo
  {author} {\bibfnamefont {Y.}~\bibnamefont {Cheng}}, \bibinfo {author}
  {\bibfnamefont {T.}~\bibnamefont {Zhang}}, \bibinfo {author} {\bibfnamefont
  {D.}~\bibnamefont {Wu}},\ and\ \bibinfo {author} {\bibfnamefont
  {X.}~\bibnamefont {Liu}},\ }\bibfield  {title} {\bibinfo {title} {Topological
  {{Creation}} of {{Acoustic Pseudospin Multipoles}} in a {{Flow}}-{{Free
  Symmetry}}-{{Broken Metamaterial Lattice}}},\ }\href
  {https://doi.org/10.1103/PhysRevLett.118.084303} {\bibfield  {journal}
  {\bibinfo  {journal} {Physical Review Letters}\ }\textbf {\bibinfo {volume}
  {118}},\ \bibinfo {pages} {084303} (\bibinfo {year}
  {2017}{\natexlab{a}})}\BibitemShut {NoStop}%
\bibitem [{\citenamefont {Zhang}\ \emph
  {et~al.}(2017{\natexlab{b}})\citenamefont {Zhang}, \citenamefont {Tian},
  \citenamefont {Cheng}, \citenamefont {Liu},\ and\ \citenamefont
  {Christensen}}]{ZhangPhysRevB2017}%
  \BibitemOpen
  \bibfield  {author} {\bibinfo {author} {\bibfnamefont {Z.}~\bibnamefont
  {Zhang}}, \bibinfo {author} {\bibfnamefont {Y.}~\bibnamefont {Tian}},
  \bibinfo {author} {\bibfnamefont {Y.}~\bibnamefont {Cheng}}, \bibinfo
  {author} {\bibfnamefont {X.}~\bibnamefont {Liu}},\ and\ \bibinfo {author}
  {\bibfnamefont {J.}~\bibnamefont {Christensen}},\ }\bibfield  {title}
  {\bibinfo {title} {Experimental verification of acoustic pseudospin
  multipoles in a symmetry-broken snowflakelike topological insulator},\ }\href
  {https://doi.org/10.1103/PhysRevB.96.241306} {\bibfield  {journal} {\bibinfo
  {journal} {Physical Review B}\ }\textbf {\bibinfo {volume} {96}},\ \bibinfo
  {pages} {241306(R)} (\bibinfo {year} {2017}{\natexlab{b}})}\BibitemShut
  {NoStop}%
\bibitem [{\citenamefont {Bernevig}\ \emph {et~al.}(2006)\citenamefont
  {Bernevig}, \citenamefont {Hughes},\ and\ \citenamefont
  {Zhang}}]{BernevigScience2006}%
  \BibitemOpen
  \bibfield  {author} {\bibinfo {author} {\bibfnamefont {B.~A.}\ \bibnamefont
  {Bernevig}}, \bibinfo {author} {\bibfnamefont {T.~L.}\ \bibnamefont
  {Hughes}},\ and\ \bibinfo {author} {\bibfnamefont {S.-C.}\ \bibnamefont
  {Zhang}},\ }\bibfield  {title} {\bibinfo {title} {Quantum {{Spin Hall
  Effect}} and {{Topological Phase Transition}} in {{HgTe Quantum Wells}}},\
  }\href {https://doi.org/10.1126/science.1133734} {\bibfield  {journal}
  {\bibinfo  {journal} {Science}\ }\textbf {\bibinfo {volume} {314}},\ \bibinfo
  {pages} {1757} (\bibinfo {year} {2006})}\BibitemShut {NoStop}%
\bibitem [{\citenamefont {Deng}\ \emph {et~al.}(2017)\citenamefont {Deng},
  \citenamefont {Ge}, \citenamefont {Tian}, \citenamefont {Lu},\ and\
  \citenamefont {Jing}}]{DengPhysRevB2017}%
  \BibitemOpen
  \bibfield  {author} {\bibinfo {author} {\bibfnamefont {Y.}~\bibnamefont
  {Deng}}, \bibinfo {author} {\bibfnamefont {H.}~\bibnamefont {Ge}}, \bibinfo
  {author} {\bibfnamefont {Y.}~\bibnamefont {Tian}}, \bibinfo {author}
  {\bibfnamefont {M.}~\bibnamefont {Lu}},\ and\ \bibinfo {author}
  {\bibfnamefont {Y.}~\bibnamefont {Jing}},\ }\bibfield  {title} {\bibinfo
  {title} {Observation of zone folding induced acoustic topological insulators
  and the role of spin-mixing defects},\ }\href
  {https://doi.org/10.1103/PhysRevB.96.184305} {\bibfield  {journal} {\bibinfo
  {journal} {Physical Review B}\ }\textbf {\bibinfo {volume} {96}},\ \bibinfo
  {pages} {184305} (\bibinfo {year} {2017})}\BibitemShut {NoStop}%
\end{thebibliography}%

\end{document}


\title{{\Large Supplementary Information} \\ \vspace{0.20cm} Topological Phononic Logic}
\author{Harris Pirie}
\author{Shuvom Sadhuka}
\author{Jennifer Wang}
\author{Radu Andrei}
\author{Jennifer E. Hoffman}
\date{\today}

\maketitle


\begin{figure}[t]
	\includegraphics[width=0.48\textwidth]{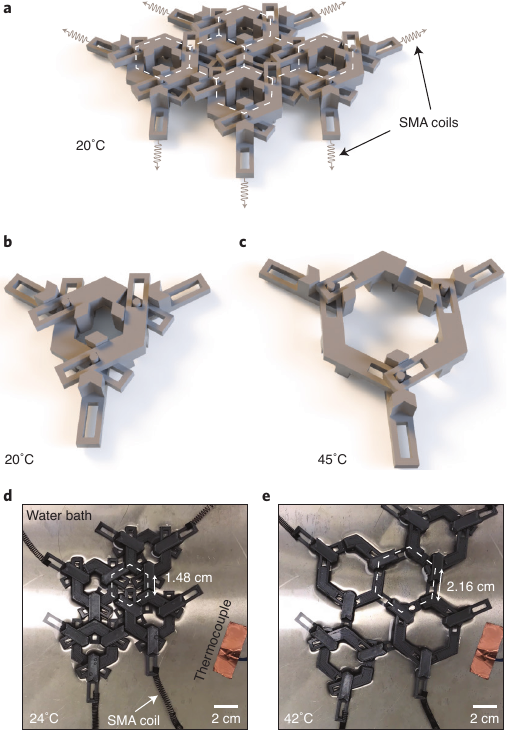}
	\caption{\label{fig:s3}
	{\bf Honeycomb lattice with a high thermal expansion coefficient.}
    ({\bf a}) Model of a honeycomb lattice that can be rapidly expanded using SMA coils to create a high thermal expansion coefficient. 
    ({\bf b}) Each unit cell is compressed below the coil's transition temperature.
   ({\bf c}) But above the transition temperature, the coils shrink dramatically to pull the unit cell open.  
    ({\bf d-e}) To experimentally measure thermal expansion, we placed the 3D-printed expandable lattice in a water bath, which we gradually heated. The lattice expanded by 46\% during a 18$\degree$C temperature increase, resulting in an effective thermal expansion coefficient of $26 \times 10^{-3}$ K$^{-1}$ (see supplementary video). 
    }
\end{figure}

\begin{figure}
	\includegraphics[width=0.48\textwidth]{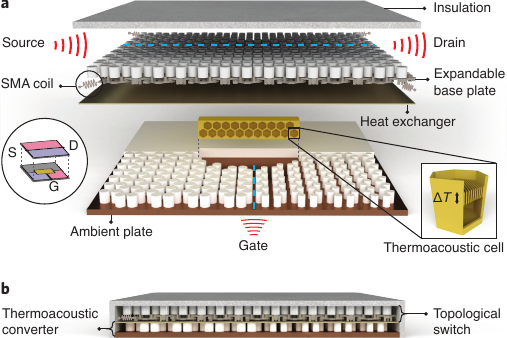}
	\caption{\label{fig:s8}
	{\bf Thermoacoustic converter.}
	 To turn `on' a topological switch, ultrasound arriving at the `gate' input needs to generate enough heat to actuate the SMA coils that expand the base plate, and to adjust the medium's speed of sound accordingly. Our design uses a separate topological waveguide (white pillars) to focus this incoming ultrasound onto an array of thermoacoustic heat pumps (yellow). Each cell in the thermoacoustic array contains a high-surface-area stack that exchanges heat with the acoustic wave to generate a temperature gradient $\Delta T$. The hot side of each stack is coupled to the expandable topological switch above (gray pillars) through a metallic heat exchanger, while the cold side is sunk to ambient temperature via a plate below. The switch is thermally insulated on all other sides, which also provides a controllable means of cooling to turn it `off'. ({\bf a}) Expanded and ({\bf b}) collapsed view of an ultrasound-controlled topological switch. 
    }
\end{figure}

\section{Engineering a high thermal expansion coefficient}
\mypara
To function correctly, our topological switch requires a base plate with an unusually high thermal expansion coefficient of $1.61 \times 10^{-3}$ K$^{-1}$. Here we show that it is possible to greatly exceed this value over a limited temperature range with  a shape-memory alloy (SMA), such as nitinol. SMAs make excellent thermal actuators because they can be trained to remember different shapes at different temperatures. For example, commercially available SMA coils can have one length when cooled and a different length when heated. The coil's expansion scales with its overall length, allowing virtually any thermal expansion coefficient to be achieved. To illustrate this potential, we built an expandable honeycomb lattice and actuated it with inch-long Nitinol springs that have a transition temperature of 45$\degree$C (see Fig.~\ref{fig:s3}{(a-c)}). We placed the lattice in a water bath to ensure an even temperature distribution while heating. As we increased the bath temperature from 24$\degree$C to 42$\degree$C, the lattice constant expanded from 1.48 cm to 2.16 cm (Fig.~\ref{fig:s3}{(d-e)} and supplementary video), corresponding to a thermal expansion coefficient of $26 \times 10^{-3}$ K$^{-1}$, a factor of 16 greater than that required for our topological phononic logic design.

\section{Thermoacoustic converter}
\mypara
For the output of one switch to toggle the state of another, we require a device to convert ultrasonic energy into heat. We employ a thermoacoustic heat pump---a resonant chamber containing a high-surface-area stack (see right inset in Fig.~\ref{fig:s8}(a)). During each acoustic cycle, the gas within such a chamber expands and contracts, exchanging heat with the stack to generate a temperature gradient across it. A thermoacoustic device can act as a heat pump or a refrigerator depending on how the stack is sunk to ambient temperature. Ultrasonic thermoacoustic devices have demonstrated heat transfer up to 1 W and coefficient of performance of 2-3 in cooling mode \cite{SAK04, CTM21}; in heating mode they will operate with even greater efficiency. Further improvements in performance could be reached by increasing the Q factor of the resonator or by raising the gas pressure. Our proposed heating stage, shown in Fig.~\ref{fig:s8}, uses a topological waveguide (white pillars) to focus incoming ultrasound towards an array of these thermoacoustic heat pumps (copper block). We estimate that a 50 W ultrasound transducer would provide sufficient heat to turn `on' a switch like the one in Fig.~4 in a few seconds. When the control signal is absent, the switch will turn `off' in a few seconds via natural cooling controlled by the thickness of its thermal insulation. A 5-mm-thick sheet of polystyrene insulation provides adequate cooling to balance the incoming heat at 90$\degree$C, but the exact thickness should be tuned to account for uncertainties in the incoming acoustic power, the thermoacoustic converter efficiency, the convective heat transfer, and neglected losses.

\begin{figure}[t]
	\includegraphics[width=0.48\textwidth]{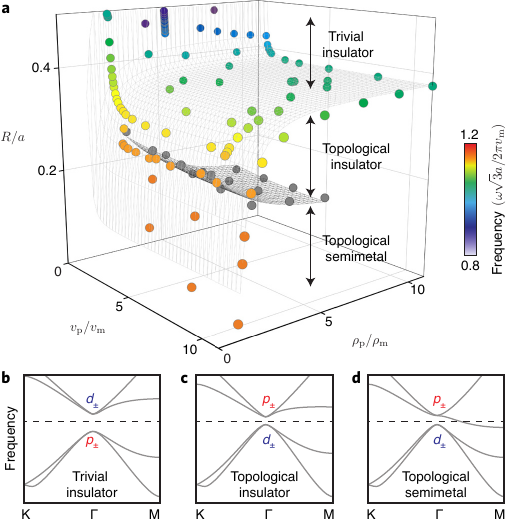}
	\caption{\label{fig:s4}
	{\bf Restrictions on topological phase space.}
	({\bf a}) The phase space for an acoustic honeycomb metamaterial is separated into a region with normally ordered $p$ and $d$ modes at the $\Gamma$ point (above the colored points), and a region where these modes are inverted (below the colored points). The latter region is further divided based on whether there is a complete or partial spectral gap between $p$ and $d$ modes (gray points).
	({\bf b-d}) Schematic band structures for each region of topological phase space.
    }
\end{figure}

\section{Material selection}
\mypara
The choice of materials for a topological waveguide must satisfy three conditions. First, the two pillar/medium combinations that form the waveguide must appear on opposite sides of the topological phase boundary in Fig.~2(c). Second, they must have overlapping spectral gaps at the $\Gamma$ point. Third, there should be no intra-gap states elsewhere in the Brillouin zone. The third condition begins to fail for small $\tilde{r}$ due to $M$-point eigenmodes that infringe upon the gap, producing a band structure similar to that of a semimetal (see Fig.~\ref{fig:s4}{(d)}). Consequently, a functioning waveguide is further restricted to configurations above the ($\tilde{v}$, $\tilde{\rho}$, $\tilde{r}$) surface where this semimetal behavior first begins (gray points in Fig.~\ref{fig:s4}{(a)})). In practice, this restriction is nearly always satisfied by using pillars with a large radius of $R \gtrsim 0.2a$. Thus a large volume of parameter space remains viable for a successful design; several candidate materials are listed in Table \hyperref[tbl:s1]{S1}. 

\begin{table}[h]
 \caption{\label{tbl:s1} 
 A pair of pillar/medium lattices with $\tilde{r}=0.32$ and 0.42 on either side of their boundary will function as a viable topological switch if the two band structures contain overlapping but opposite-sign gaps at the $\Gamma$ point. This table lists $\Gamma$-point eigenfrequencies for the ($p_\pm$, $d_\pm$) modes and the total gap overlap ($\Delta_\mathrm{overlap}$) for several example systems, in dimensionless units $\tilde{\omega} = \omega \sqrt{3}a / 2\pi v_{\mathrm{m}}$. Combinations in black work, but those in red fail because the modes do not invert.}
 \centering
 \begin{tabular}{ll|cccc}
 \toprule
 & \multicolumn{1}{l|}{} & \multicolumn{4}{c}{Pillar material}\\
  \multicolumn{2}{l|}{\ Medium} & \multicolumn{1}{c}{Steel} & \multicolumn{1}{c}{Aluminum} & \multicolumn{1}{c}{Rubber} & \multicolumn{1}{c}{HDPE}\\
 \hline
 \parbox[t]{4.5pt}{\multirow{3}{*}{\rotatebox[origin=c]{90}{$\tilde{r}=0.32$}}} 
   		& Air	& 0.96, 1.03	& 0.96, 1.03	& 0.96, 1.03	& 0.96, 1.03		\\
   		& Diesel & 1.01, 1.05	& 1.08, 1.10	& 1.11, 1.12 	& \red{1.17, 1.20}	\\
   		& Water 	& 1.02, 1.06	& 1.10, 1.11	& 1.11, 1.13 	& \red{1.18, 1.23}	\\
 \hline
 \parbox[t]{4.5pt}{\multirow{3}{*}{\rotatebox[origin=c]{90}{$\tilde{r}=0.42$}}} 
   		& Air	& 1.00, 0.96	& 1.00, 0.96	& 1.00, 0.96	& 1.00, 0.96		\\
		& Diesel	& 1.08, 1.01	& 1.19, 1.09	& 1.16, 1.11	& \red{1.21, 1.34}	\\
   		& Water 	& 1.09, 1.02	& 1.22, 1.11  & 1.15, 1.12	& \red{1.23, 1.35}	\\  
 \hline
 \parbox[t]{4.5pt}{\multirow{3}{*}{\rotatebox[origin=c]{90}{$\Delta_\mathrm{overlap}$}}} 
   		& Air	    & 0.041	& 0.041	& 0.041	& 0.041	    \\
		& Diesel	& 0.037	& 0.005	& 0.005	& \red{-}   \\
   		& Water 	& 0.035	& 0.002	& 0.005	& \red{-}	\\

\bottomrule
 \end{tabular}
 \end{table}

\begin{figure}
	\includegraphics[width=0.5\textwidth]{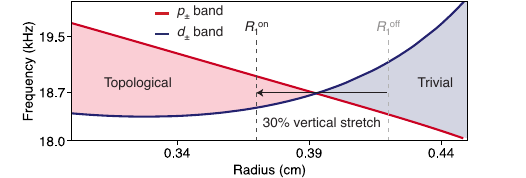}
	\caption{\label{fig:s5}
	{\bf A topological switch made from rubber.}
	The calculated $\Gamma$-point eigenvalues for a hexagonal lattice of rubber pillars in air with $a= 1$ cm contains an accidental degeneracy when the pillar radius is 0.39 cm.  Vertically stretching the pillars by 30\% toggles an initial trivial waveguide, with unstretched radii $R_1$ = 0.42 and $R_2$ = 0.48 cm, to a topological waveguide with radii $R_1$ = 0.37 and $R_2$ = 0.42 cm.
    }
\end{figure}

\mypara
Among the viable material systems, a straightforward choice for building an externally controllable topological switch like the one in Fig.~1(c) is to use rubber pillars in air. Vulcanized rubber can easily elongate by several hundred percent, providing considerable tunability in the pillar radius \cite{B50}. In addition, rubber has a Poisson's ratio $\nu$ = 0.5, implying that its density and longitudinal speed of sound do not change under tensile strain. An example design contains two sets of rubber pillars with radii $R_1 = 0.42$ cm and $R_2 = 0.48$ cm, height $H = 1$ cm, and spacing $a = 1$ cm, which each initially form a trivial insulator (`off' state). Stretching the pillars by an amount $\Delta H$, changes their radius by  
\begin{align}
	\Delta R = - R \left(1 - \left(1 + \frac{\Delta H}{H} \right)^{-\nu} \right).
\end{align}
In our design, the waveguide turns `on' when the pillars are stretched by 30\%, requiring an applied tensile stress of 0.2 MPa \cite{B50}. This stretch reduces the pillar's radii to $R_1 = 0.37$ cm and $R_2 = 0.42$ cm, and the $R_1$ pillars form a topological phase as their radius is smaller than the critical value for a rubber/air system of $R^* = 0.39$ cm (see Fig.~\ref{fig:s5}). The operational frequency of this device is 18.7 kHz, and it has a bulk spectral gap overlap of 500 Hz before and after switching.

\section{Robustness of topological states}
\mypara
Topological materials are characterized by a $\mathbb{Z}_2$ index, which predicts the presence of gapless boundary modes. A widely applied approach to calculating the $\mathbb{Z}_2$ index for photonic and phononic metamaterials is to apply $k\cdot p$ perturbation theory \cite{WuPhysRevLett2015, MeiPhysRevB2012, HeNatPhys2016, ZhangPhysRevLett2017, ZhangPhysRevB2017}.  In this framework, the effective low-momentum Hamiltonian for our specific system is similar to the BHZ model of the quantum spin Hall effect in a two-dimensional topological insulator \cite{BernevigScience2006}.  The complete calculation is presented in Ref.~\cite{HeNatPhys2016}, and predicts non-trivial spin Chern numbers only when the $p_\pm$ and $d_\pm$ bands are inverted, as expected for a topological phase.

\begin{figure}
	\includegraphics[width=0.48\textwidth]{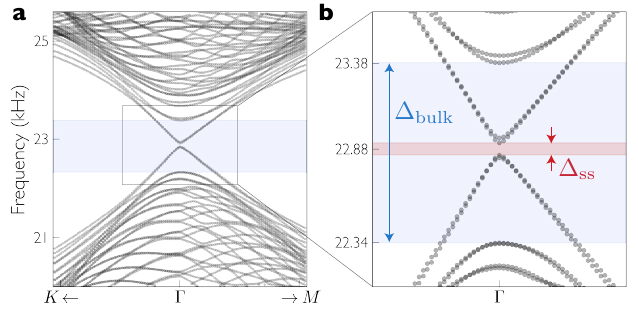}
	\caption{\label{fig:s1}
	{\bf Breaking $C_{6v}$ symmetry subtly gaps the Dirac point.}
    ({\bf a}) When `on', a steel/air topological switch like the one in Fig.~4(c) hosts protected boundary states that disperse linearly within the inverted bulk band gap (blue).
    ({\bf b}) These states are protected by a pseudo-time-reversal symmetry constructed from geometric rotations in the $C_{6v}$ point group.  At the interface, this symmetry is broken, leading to a small unwanted gap in the Dirac spectrum (red). The interface geometry can be optimized to reduce $\Delta_{\mathrm{ss}}$ with respect to $\Delta_{\mathrm{bulk}}$.
    }
\end{figure}

\mypara
Our logic gate exploits robust edge modes, protected by a pseudo-time-reversal symmetry built from discrete lattice rotations in the $C_{6v}$ group \cite{WuPhysRevLett2015}. However, $C_{6v}$ symmetry is broken at the interface between two different lattices, allowing states of opposite pseudospin to mix, and opening a small gap around the Dirac point proportional to the degree of symmetry breaking.  In our system, this mini-gap is about 7\% of the bulk gap (see Fig.~\ref{fig:s1}).  This issue was previously explored by He {\it et al.}~\cite{HeNatPhys2016}, who significantly reduced the mini-gap to less than 0.02\% of the bulk gap by introducing an additional row of pillars at the interface. Importantly for us, they demonstrated that the mini-gap can be optimized {\it after} selecting material parameters because it is a property of the interface, not the bulk.  Improvements beyond their benchmark are possible, and designing an interface that most effectively reduces the mini-gap is an important area of further work.

\begin{figure}[h]
	\includegraphics[width=0.48\textwidth]{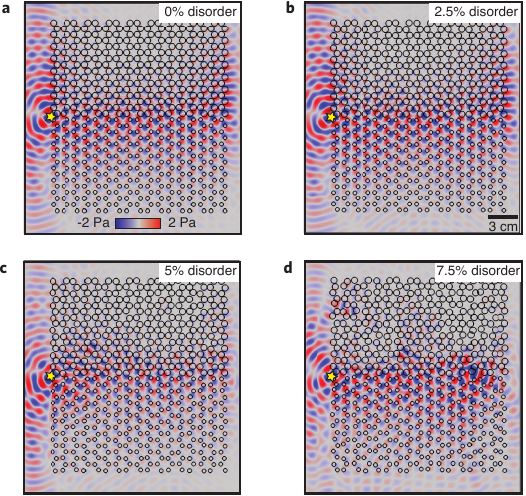}
	\caption{\label{fig:s6}
	{\bf Edge states are largely immune to fabrication disorder.}
    ({\bf a}) This simulation of a $20 \times  20$ cm$^2$ topological waveguide contains 544 steel pillars in air ($R_1=1.9$ mm, $R_2 = 3.1$ mm, $a=7.0$ mm). A 27.5 kHz point source (yellow star) excites waves that are localized to the interface between small and large pillars.
    ({\bf b-d}) The edge states remain fairly localized even if the pillars are allowed to move from their nominal positions by up to 7.5\%, suggesting the metamaterial is robust to moderate fabrication errors.
    }
\end{figure}

\begin{figure}[h]
	\includegraphics[width=0.49\textwidth]{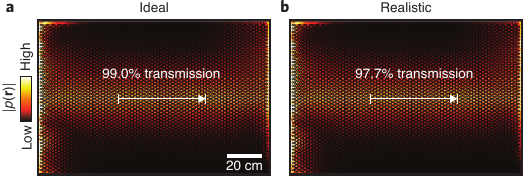}
	\caption{\label{fig:s7}
	{\bf Topological waveguides are efficient.}
    ({\bf a}) Unless otherwise stated, our topological switch is simulated using ideal materials that do not incur inherent losses due to dissipation. It has a high transmission efficiency of 99\% over a long 50 cm section (arrow), demonstrating the absence of significant backscattering. 
    ({\bf b}) The transmission efficiency remains high even with realistic, lossy materials that include the viscosity of 22 $\mu$Pa$\cdot$s for 90$\degree$C air and a damping coefficient of 0.001 for steel.
    }
\end{figure}

\mypara
The benefit of topological phononic logic is that transport is relatively immune to defects, leading to more efficient devices.  In our example, defects that preserve pseudo-time-reversal symmetry are unable to mix pseudospin states, eliminating backscattering. It has been shown previously that backscattering is negligible for cavities, bends, and disorder \cite{HeNatPhys2016}. On the other hand, backscattering is present for defects that explicitly break $C_{6v}$ symmetry, as they mix pseudo-spin states and open a mini-gap in the spectrum; these should be avoided \cite{DengPhysRevB2017}. In practice, the most likely defects in our device are random errors in the pillar's positions arising due to fabrication imperfections. Although these errors tend to further break the $C_{6v}$ symmetry and enhance the mini-gap in the edge spectrum, the device is surprisingly robust to them. We simulated a realistic waveguide with varying amounts of random disorder in the pillar's position (see Fig.~\ref{fig:s6}). The highest reasonable amount of disorder is about 5\% of the lattice constant---anything higher and the larger pillars begin to touch each other. The transmission through our waveguide is virtually unaffected by such moderate disorder, and it appears only mildly reduced even for unrealistically high disorder above 5\%. Other loss mechanisms are present in our device, such as energy dissipation in the pillars and air, but they have a minimal impact on its transmission (see Fig.~\ref{fig:s7}).

\begin{figure}[t]
	\includegraphics[width=0.48\textwidth]{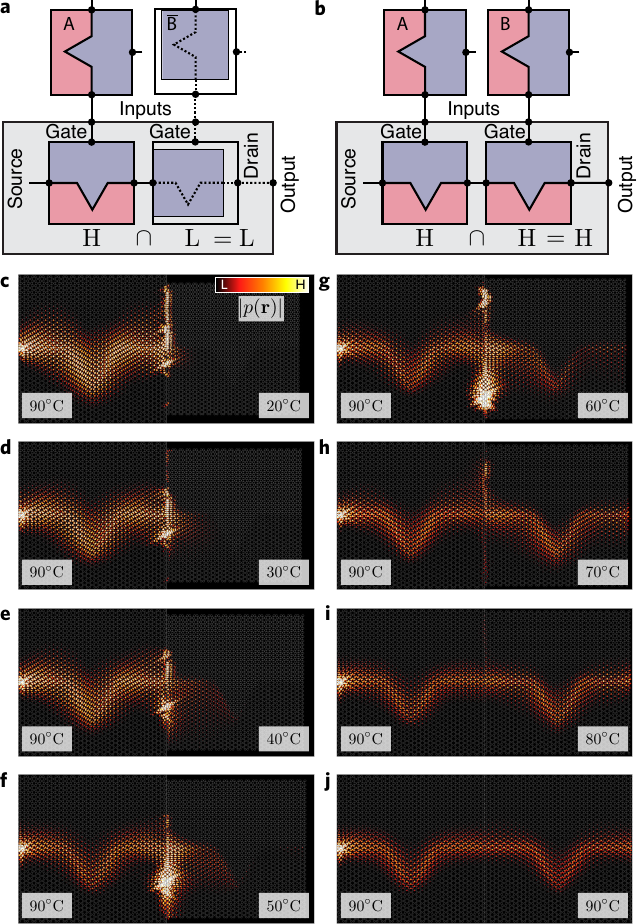}
	\caption{\label{fig:s2}
	{\bf Switching `on' a topological {\sc and} gate}
    ({\bf a-b}) When either input to a topological {\sc and} gate (gray area) is low, no signal can propagate to its output. 
    ({\bf c}) Instead, it scatters at the interface, due to an abrupt change in lattice constant, and decays exponentially into the `off' switch (right), as shown in this simulation of two connected steel/air topological waveguides, identical to those in Fig.~4(c) and (e).
   ({\bf d-f}) As the right-hand switch is heated, its band gap reduces and the signal decays more slowly.  
   ({\bf g-j}) Above 50$\degree$C, the band gap is inverted in the lower part of the second switch, yielding topological edge modes that become more localized with further heating.
    }
\end{figure}

\section{Phononic logic gates}
\mypara
A topological {\sc and} gate allows propagation only if its two control signals are high. It is created from two phononically controlled topological switches connected in series, as shown in Fig.~\ref{fig:s2}{(a-b)}. If either control signal is low, the corresponding switch will be at room temperature, with both of its sides in a trivial insulating phase. The propagating signal is stopped at the first cold switch that it meets, where it decays exponentially with a length scale inversely proportional to the size of the band gap (Fig.~\ref{fig:s2}{(c)}).  Transmission through the gate becomes noticeable only after the cold switch transitions from a trivial-trivial insulator to a topological-trivial insulator, which happens at about 50$\degree$C (Fig.~\ref{fig:s2}{(f)}).  Heating beyond this point has two effects: it increases the size of the band gap in the second switch, causing the edge states to become more localized; and it reduces the backscattering that results from the lattice constant changing abruptly at the junction. Surprisingly, the gate can maintain stable transmission despite temperatures differences of around 10$\degree$C between switches (Fig.~\ref{fig:s2}{(i)}). 

\begin{figure}
	\includegraphics[width=0.48\textwidth]{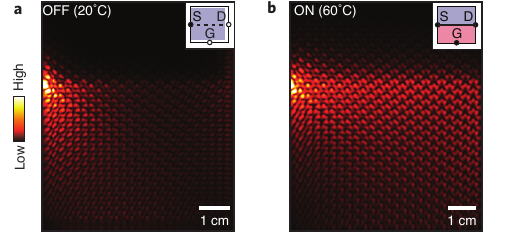}
	\caption{\label{fig:s9}
	{\bf Low-power logic gate.}
	 To reduce the power required to transition a topological switch, we designed a small steel/argon waveguide operating at 106.9 kHz with parameters $R_1=0.687$ mm, $R_2 = 0.82$ mm, $a=1.7$ mm, and thermal expansion coefficient of the underlying base plate $\alpha = 1.651\times 10^{-3}$ K$^{-1}$. This design requires about 0.8 J of heat to transition. Simulations of the ({\bf a}) `off', and ({\bf b}) `on' states of the low-power switch.
	 }
\end{figure}

\section{Faster, Lower-power logic gates} 
\mypara
A key step to lower the power consumption of our logic gates is to decease their total heat load by optimizing their size, materials, and operating temperature range:
\begin{itemize}
    \item {\bf Reduce the total size.} Our design is easily miniaturized because the governing topological physics is scale invariant. Reducing the dimensions by a factor of five is still within typical fabrication tolerances, and yields a logic gate for $\sim 100$ kHz ultrasound. The size of this device can be further reduced by removing unused pillars from the $R_2$ region, where the insulating gap is large throughout the transition (see Fig.~3(b)).  As the waveguide is a quasi-2D system, it can be as thin as possible, provided the pillar height is $\gtrsim\lambda$.  Combining these constraints, we designed a miniature switch with dimensions 6.6 cm $\times$ 7.0 cm $\times$ 0.3 cm (see simulation in Fig.~\ref{fig:s9}).    
    \item {\bf Lower the operating temperature range.} The main issue with smaller temperature ranges is that the insulating gap is reduced and hence the edge state can delocalize. However, operating between 30$\degree$C and 70$\degree$C gives almost identical edge-state localization to that at 20$\degree$C and 90$\degree$C (compare Fig.~\ref{fig:s2} (c, d) and (h, j)). To keep the `off' temperature at 20$\degree$C, we can adjust the pillar radius $R_1$, which changes the filling ratio at each temperature $\tilde{r} = R_1/a(T)$, where $a(T)$ is the temperature-dependent pillar spacing. Specifically, our miniature waveguide uses $R_1$ = 0.687 mm to create equal but inverted spectral gaps at 20$\degree$C and 60$\degree$C (see Fig.~\ref{fig:s9}). This smaller operating temperature range of $\Delta T=40$ K requires a slightly higher thermal expansion coefficient of 1.651$\times 10^{-3}$ K$^{-1}$, compared to 1.614$\times 10^{-3}$ K$^{-1}$ in our original design.    
    \item {\bf Select materials with good thermal characteristics.} To toggle a switch, we require only the gas medium and the SMA coils to reach the transition temperature, whereas the pillars and expandable base plate can be thermally insulated from the temperature-sensitive components. We can further reduce the heat load by selecting a working gas with a low specific heat, such as argon. 
\end{itemize}
With these collective improvements, we estimate that it will take only 0.8 J of heat to actuate a miniature argon-based switch with commercially available SMA micro-springs.


\bibliography{refs}